\documentclass[aps,prd,10pt,amsmath,floats,floatfix,twocolumn,altaffilletter,nofootinbib,shownopacs,longbibliography]{revtex4-2}
\usepackage[T1]{fontenc}
\usepackage[utf8]{inputenc}
\usepackage{lmodern}
\usepackage{verbatim}
\usepackage{physics}
\usepackage[dvipsnames,usenames]{xcolor}
\definecolor{linkcolor}{rgb}{0.6,0.0,0.0}
\definecolor{applegreen}{rgb}{0.55, 0.71, 0.0}
\usepackage{appendix}
\usepackage{booktabs}
\usepackage{enumitem}
\usepackage[hypertexnames=false, unicode, colorlinks=true, linkcolor=linkcolor,
citecolor=linkcolor, filecolor=linkcolor,urlcolor=linkcolor,
pdfusetitle]{hyperref}

\usepackage[all]{hypcap}
\usepackage{graphicx}
\usepackage{color}
\usepackage{xspace}
\usepackage{amssymb}
\usepackage[normalem]{ulem} 
\usepackage{bm} 
\usepackage{orcidlink}
\usepackage{microtype}
\usepackage[english]{babel}
\usepackage{blindtext}

\graphicspath{%
  {figs/}%
}

\def\beq{\begin{equation}\begin{aligned}}
\def\eeq{\end{aligned}\end{equation}}

\newcommand{\PP}{\vec{\mathbb{P}}}    
\newcommand{\bS}{\vec{\mathbf{S}}}    
\newcommand{\hZ}{\hat{\mathbf{z}}}    
\newcommand{\Lhat}{\widehat{\mathbf{L}}} 
\newcommand{\Io}{\mathcal{I}_0}
\newcommand{\Ii}{\mathcal{I}_1}

\begin{document}

\newcommand{\TIFR}{\affiliation{Tata Institute of Fundamental Research, Homi Bhabha Road, Colaba, Mumbai 400005, India}}

\title{Stability of Collective Neutrino Oscillations -- A Distributional Approach}

\author{Rupak Majumder\,\orcidlink{0009-0003-9275-8571}\,}
\email{rupak.majumder@tifr.res.in}

\author{Dwaipayan Mukherjee\,\orcidlink{0009-0005-2204-2249}\,}
\email{dwaipayan.mukherjee@tifr.res.in}

\author{Shamik Gupta\,\orcidlink{0000-0002-6080-4890}\,}
\email{shamik.gupta@theory.tifr.res.in}

\author{Basudeb Dasgupta\,\orcidlink{0000-0001-6714-0014}\,}
\email{bdasgupta@theory.tifr.res.in}
\TIFR{}

\date{September 3, 2026;~{Preprint No.\,TIFR/TH/26-23}}

\begin{abstract}

We study the stability of collective neutrino oscillations using a
distributional approach motivated by the statistical mechanics of Kuramoto synchronization. Treating the ensemble of neutrino flavor polarization vectors
in the thermodynamic limit $N\to\infty$, we derive an exact nonlinear
Fokker--Planck (continuity) equation for the one-body distribution
$F(\bS,\omega,t)$ on the flavor sphere.  This equation admits a two-parameter
family of azimuthally symmetric stationary solutions, whose stability we
analyze by linearizing around them.  The resulting eigenvalue condition
determines the growth or decay rate of small perturbations from
\emph{any} initial distribution -- not merely from a state close to full
flavor coherence -- thereby going significantly beyond the conventional
linear stability analysis of collective modes.  In special limits the
condition reproduces known synchronization thresholds in the
two-beam model, providing a non-trivial check of the framework.  We
present analytical results for the eigenvalue equation and explore
stability phase diagrams for physically relevant frequency distributions.
\end{abstract}

\maketitle


\tableofcontents

\section{Introduction}
\label{sec:intro}

A core-collapse supernova offers an extreme laboratory for neutrino oscillations. Most notably, neutrinos in the dense post-bounce environment experience mutual forward-scattering with a potential $\mu = \sqrt{2}\,G_F n_\nu$ that can far exceed the vacuum oscillation frequency $\omega = \Delta m^2/(2E)$~\cite{Pantaleone:1992eq, Sigl:1992fn}. This $\nu$--$\nu$ interaction drives \emph{collective flavor oscillations} that can lead to sizable flavor conversion on dynamically relevant timescales~\cite{Duan:2010bg, Chakraborty:2016yeg, Mirizzi:2015eza, Tamborra:2020cul, Volpe:2023met, Johns:2025mlm}.

A central question in collective neutrino oscillations is whether a given flavor configuration is dynamically unstable. Such instabilities can be broadly classified based on the characteristic time scale: \emph{Slow} flavor instabilities develop on a frequency scale $\sim\sqrt{\omega\mu}$~\cite{Duan:2005cp,Hannestad:2006nj,Fiorillo:2024pns}; they are associated with crossings in the energy spectra of different flavors~\cite{Dasgupta:2009mg}. \emph{Fast} flavor instabilities~\cite{Sawyer:2005jk,Sawyer:2008zs,Fiorillo:2024b} can develop on frequencies $\sim\mu \gg \sqrt{\omega\mu}$,  and may occur generically wherever the electron lepton number flux has an angular crossing~\cite{Izaguirre:2016gsx,Capozzi:2017gqd}. More generally, the occurrence of collective instabilities  is intimately connected to the existence of crossings in the flavor distributions~\cite{Morinaga:2021vmc,Dasgupta:2021gfs,Dasgupta:2025quc}.

The standard tool for assessing whether a given neutrino configuration will undergo collective instability is \emph{linear stability analysis} (LSA)~\cite{Izaguirre:2016gsx,Capozzi:2017gqd,Banerjee:2011fj,Airen:2018nvp}. One linearizes the equations of motion (EoMs) for the polarization vectors around an initial state close to full flavor coherence -- with all neutrinos close to being flavor eigenstates -- and examines whether small perturbations grow exponentially. The growth rate of the fastest-growing Fourier mode then sets the timescale for the onset of conversions.  While powerful and widely used, this approach has an intrinsic limitation: it speaks only to the stability of one specific initial condition.  Whether partially converted or more broadly distributed flavor configurations are themselves stable is a question that standard LSA does not address. In particular, after an instability has partially reshaped the ensemble, can the resulting distribution remain stable, or can it undergo further collective evolution? Similarly, can different distributions of polarization vectors over the flavor sphere with the same frequency spectrum exhibit qualitatively different stability properties?

In this paper, we fill this gap by developing a \emph{distributional} stability analysis. Our approach is inspired by the kinetic theory of coupled oscillators, in particular, the Kuramoto model~\cite{kuramoto1984chemical, strogatz2000kuramoto, acebron2005kuramoto, gupta2014kuramoto}.  In the thermodynamic limit $N\to\infty$, the collective neutrino system is described by a one-body distribution $F(\bS,\omega,t)$ on the two-sphere $S^2$, obeying an exact nonlinear Fokker--Planck (continuity) equation. This equation admits a rich family of stationary solutions, and we can systematically ask about the stability of \emph{any} such solution by linearizing the kinetic equation around it.  The resulting eigenvalue condition -- a Fredholm-type integral equation for the perturbation growth rate $\Omega$ -- determines whether the state is stable or unstable, regardless of its degree of polarization. This differs from the usual matrix-valued quantum-kinetic description, in which flavor is encoded in occupation-number density matrices.

As an application of our formalism, we derive a general stability criterion for the class of azimuthally symmetric stationary states, which we call the ``north-south'' family.  This family is parameterized by two numbers: $\kappa$, which controls the degree of polar concentration, and $\alpha$, which controls the asymmetry between the neutrino and antineutrino populations. The uniform state (all modes spread uniformly over $S^2$) and the two back-to-back beam configuration are the two limiting cases, $\kappa=0$ and $\kappa\to\infty$ respectively.  In the latter limit, our eigenvalue condition reproduces the known synchronization threshold of the two-beam model~\cite{Hannestad:2006nj}, providing a sanity check.  
Beyond this limit, the same condition determines both boundaries of the bipolar instability and extends them to finite-$\kappa$ north-south states and continuous frequency spectra, for which the unstable interval can narrow or disappear altogether. In the examples, the $z-$projection of the total polarization vectors provides a diagnostic of the bipolar instability, while a late-time transverse order parameter distinguishes vacuum-like dephasing from synchronized motion within the linearly stable region \cite{Raffelt:2010za}.

The remainder of this paper is organized as follows.  In Sec.~\ref{sec:formalism} we introduce the model and derive the Fokker--Planck equation for the distribution $F(\bS,\omega,t)$.  The stationary solutions and their basic properties are presented in Sec.~\ref{sec:stationary}.  The linearized stability analysis and the eigenvalue condition are developed in Sec.~\ref{sec:stability}.  We discuss analytical results and special limiting cases in Sec.~\ref{sec:analytic}, and explore the stability phase diagram numerically in Sec.~\ref{sec:numerical}.  We conclude with a summary and outlook in Sec.~\ref{sec:discussion}.  Technical derivations are collected in the Appendices.

\section{Formalism}
\label{sec:formalism}

\subsection{Equations of Motion}
\label{sec:eom}

We consider $N$ two-flavor neutrino modes, each described by a unit
three-vector (Bloch/polarization vector) $\bS_j \in S^2$,
$j = 1,\ldots,N$.  The vacuum precession frequencies $\omega_j$ are
drawn from a distribution that encodes the neutrino energy
spectrum.  In a spatially homogeneous, isotropic geometry with
neutrino--neutrino potential $\mu = \sqrt{2}\,G_F n_\nu$, the EoMs read
\begin{equation}
  \frac{d\bS_j}{dt}
  = \omega_j\,\hZ \times \bS_j
  + \frac{\mu}{N}\sum_{k=1}^{N}\bS_k \times \bS_j.
  \label{eq:EoM}
\end{equation}
Here $\hZ$ is the mass axis in the Bloch representation, aligned with
the mass basis.  The first term generates the individual-mode
precession; the second is the collective neutrino--neutrino interaction,
which we simplify to have no angular ($1-\mathbf{v}\cdot\mathbf{v}'$)
weighting (i.e., the \emph{single-angle} approximation).  

The full \emph{multi-angle} dependence can be restored by assigning a velocity $\mathbf v_j$ to each mode and reverting to the angular kernel,
\begin{equation}
  \frac{d\bS_j}{dt} = \omega_j\,\hZ \times \bS_j
  + \frac{\mu}{N}\sum_{k=1}^{N}(1-\mathbf v_j\cdot\mathbf v_k)\,\bS_k \times \bS_j.
  \label{eq:EoM_multi_angle}
\end{equation}
For an isotropic angular distribution, the term proportional to $\mathbf v_j\cdot\mathbf v_k$ vanishes upon angular averaging, yielding Eq.~\eqref{eq:EoM}. Generally, the relative angle $\cos\theta_{jk}=\mathbf v_j\cdot\mathbf v_k$ makes the coupling direction-dependent, so that modes on different trajectories feel a mean field labeled by ${\bf v}_j$ rather than a single one. We work throughout in the single-angle approximation,
Eq.~\eqref{eq:EoM}, as it provides the natural point of comparison with existing work on collective oscillations and Kuramoto-type studies. 

In the case of direction-dependent coupling, one may regard the set $\{\mu_{jk}\}$, with $\mu_{jk}$ denoting the coupling between the $j$th and $k$th modes, as quenched-disordered random variables drawn from a prescribed probability distribution. For a given disorder realization, one can group pairs of modes with the same coupling and proceed with the analysis presented in this work, followed by an average over disorder realizations. However, this averaging is nontrivial: quenched disorder can lead to strong sample-to-sample fluctuations and non-self-averaging behavior, whereby different realizations may exhibit qualitatively distinct collective behavior even as $N \to \infty$~\cite{PhysRevLett.77.3700,PhysRevE.52.3469}. Moreover, disorder averaging cannot generally be replaced by using the mean coupling. Thus, the nature of the resulting behavior and the extent of self-averaging can only be established through a detailed analysis of the quenched-disordered system. Given the technical intricacies involved, we leave this investigation for future.

For our system, the collective, or synchronization, vector
\begin{equation}
  \PP = \frac{1}{N}\sum_{j=1}^{N}\bS_j
  \label{eq:OP}
\end{equation}
plays the role of an order parameter: $|\PP| = 1$ corresponds to full polarization, while $|\PP| = 0$ signals complete depolarization.
In terms of $\PP$, Eq.~\eqref{eq:EoM} becomes
\begin{equation}
  \frac{d\bS_j}{dt}
  = \omega_j\,\hZ \times \bS_j
  + \mu\,\PP \times \bS_j,
  \label{eq:EoM_OP}
\end{equation}
making manifest the ``mean-field'' structure of the interaction.

\subsection{Continuity Equation}
\label{sec:FP}

In the thermodynamic limit $N\to\infty$, the discrete sum over modes is
replaced by a continuous distribution.  We introduce the one-body
density $F(\bS,\omega,t)$, normalized as
\begin{equation}
  \int d\omega\int_{S^2} F(\bS,\omega,t)\,d\Omega = 1,
  \label{eq:normalization_F}
\end{equation}
where $d\Omega = \sin\theta\,d\theta\,d\phi$ is the solid-angle element
on $S^2$.  The order parameter \eqref{eq:OP} becomes
\begin{equation}
  \PP = \int d\omega \int \bS\,F(\bS,\omega,t)\,d\Omega.
  \label{eq:OP_F}
\end{equation}

Following standard procedures for differential equations on
manifolds (adapted from the Kuramoto literature~\cite{kuramoto1984chemical, strogatz2000kuramoto, acebron2005kuramoto, gupta2014kuramoto, gupta2018statistical, majumder2025finitesizefluctuationsstochasticcoupled}), and extending the formalism to $S^2$~\cite{chandra2019continuous, majumder2026synchronizationannealeddisorderhigherharmonic}, one derives the continuity equation for the time evolution of $F$:
\begin{equation}
  \frac{\partial F}{\partial t}
  = -\Bigl[\omega\,\hZ\times\bS + \mu\,\PP\times\bS\Bigr]
    \cdot\nabla_{\!S} F,
  \label{eq:FP}
\end{equation}
where $\nabla_{\!S}$ is the gradient on $S^2$. Physically, the equation states that the local rate of change of the distribution is exactly balanced by the divergence of the probability current. This is an exact,
\emph{nonlinear} integro-partial differential equation because $\PP$
depends self-consistently on $F$ through Eq.~\eqref{eq:OP_F}.

\section{Stationary States}
\label{sec:stationary}

\subsection{Fully Depolarized Distribution}
\label{sec:incoherent}

The simplest stationary solution of Eq.~\eqref{eq:FP} is the
\emph{fully depolarized state},
\begin{equation}
  F_0(\bS,\omega) = \frac{g(\omega)}{4\pi},
  \label{eq:F0_uniform}
\end{equation}
for which $\PP = \mathbf{0}$, where $g(\omega)$ is any normalized frequency distribution. As $\nabla_{\!S} F_0=0$, one immediately verifies that
$\partial_t F_0 = 0$.  This state corresponds to all neutrino modes
being uniformly spread over the flavor sphere with no net flavor polarization.

\subsection{North-South Family}
\label{sec:NS_states}

A physically richer family of stationary states is
\begin{align}
    &F_0(\bS,\omega) \nonumber\\
    &= \frac{1}{4\pi}\frac{\kappa}{\sinh \kappa}
  \Bigl[\beta\,g_1(\omega)\,e^{\kappa\cos\theta}
        + (1-\beta)\,g_2(\omega)\,e^{-\kappa\cos\theta}\Bigr],
  \label{eq:F0_NS}
\end{align}
where $\theta$ is the polar angle on $S^2$, $\kappa\geq 0$ is a concentration
parameter, and $\beta\in[0,1]$ weights the relative population of modes
concentrated near the north pole (mass eigenstate $\nu_1$, $\theta=0$) versus
the south pole (mass eigenstate $\nu_2$, $\theta=\pi$).  The two separate
frequency distributions $g_1(\omega)$ and $g_2(\omega)$ (normalized to
unity) allow for different spectral shapes of the two populations,
as appropriate for neutrinos and antineutrinos.

One verifies that Eq.~\eqref{eq:F0_NS} is normalized on
$S^2\times\mathbb{R}$ through Eq.~\eqref{eq:normalization_F}.  The two limiting cases are particularly
illuminating.  As $\kappa\to 0$, the distribution approaches the uniform
incoherent state with an effective frequency distribution
$\beta g_1(\omega)+(1-\beta)g_2(\omega)$.  As $\kappa\to\infty$, the
distribution concentrates on two antipodal points,
\begin{align}
    &F_0(\bS,\omega) \xrightarrow{\kappa\to\infty}  \nonumber\\
    &
  \frac{1}{2\pi}\Bigl[\beta\,g_1(\omega)\,\delta(\cos\theta-1)
                     + (1-\beta)\,g_2(\omega)\,\delta(\cos\theta+1)\Bigr],
\end{align}
which is the distribution corresponding to two back-to-back beams, a configuration closely related to the bipolar configurations~\cite{Duan:2007mv}, also arising in the context of fast flavor conversions~\cite{Capozzi:2017gqd,Chakraborty:2016lct}.

\subsubsection{Order Parameter and Self-Consistency}
\label{sec:self_consistency}

The order parameter corresponding to the stationary state
\eqref{eq:F0_NS} can be computed using the expansion
$e^{\kappa\cos\theta} = \sum_{l=0}^\infty \sqrt{4\pi(2l+1)}\,\mathcal{I}_l(\kappa)\,Y_l^0$,
where $\mathcal{I}_l(\kappa) = i^l j_l(-i\kappa)$ are modified spherical Bessel
functions and $Y_l^m$ are spherical harmonics with $j_l$ being the usual Bessel functions.  Since $\bS$ is built from $l=1$ spherical harmonics, only
the $l=1$ component contributes, and one finds (see Appendix~\ref{app: derivation})
\begin{equation}
  \PP_0 = \Bigl(2\beta - 1\Bigr)\frac{\Ii(\kappa)}{\Io(\kappa)}\,\hZ
  \label{eq:P0}
\end{equation}
where $\Io(\kappa) = \sinh(\kappa)/\kappa$ is the modified spherical Bessel function of
order zero. For $\beta=1/2$ (equal populations at north and south poles) or $\kappa=0$
(uniform distribution), one has $\PP_0 = \mathbf{0}$.  As $\kappa\to\infty$, we have 
$\Ii(\kappa)/\Io(\kappa)\to 1$, and the order parameter saturates at
$|\PP_0| = |2\beta - 1|$.

Now, $F_0$ in Eq.~\eqref{eq:F0_NS} is indeed a stationary solution of Eq.~\eqref{eq:FP}:
since $F_0$ depends only on $\cos\theta$, one has
$\nabla_{\!S}F_0 \propto \hat\theta$, and $(\hZ\times\bS)\cdot\hat\theta = 0$
as well as $(\PP_0\times\bS)\cdot\hat\theta = 0$ because $\PP_0\parallel\hZ$ (see Appendix~\ref{app: proof stationarity} for details). 
\section{Distributional Stability Analysis}
\label{sec:stability}

\subsection{Evolution Equation for Perturbations}
\label{sec:linearization}

We now ask whether the stationary states \eqref{eq:F0_NS} are stable
with respect to perturbations.  Writing
\begin{equation}
  F(\bS,\omega,t) = F_0(\bS,\omega) + \eta(\bS,\omega,t),
  \label{eq:Fpert}
\end{equation}
and inserting  into Eq.~\eqref{eq:FP}, we obtain 
\begin{align}
  \frac{\partial\eta}{\partial t}
  &=  -\omega\bigl[\hZ\times\bS\bigr]\cdot\nabla_{\!S}\eta - \mu\bigl[\PP_0\times\bS\bigr]\cdot\nabla_{\!S}\eta \nonumber\\
  &\quad - \mu\bigl[\PP_\eta\times\bS\bigr]\cdot\nabla_{\!S}F_0-\mu[\PP_\eta\times\bS]\cdot\nabla_{\!S}\eta,
  \label{eq:eta_evolution}
\end{align}
where $\PP_\eta = \int d\omega\int\bS\,\eta(\bS,\omega,t)\,d\Omega$
is the contribution of the perturbation to the order parameter.  

\subsection{Spherical Harmonic Expansion}
\label{sec:SH_expansion}

Expanding the perturbation in spherical harmonics,
\begin{equation}
  \eta(\theta,\phi,\omega,t)
  = \sum_{l=1}^{\infty}\sum_{m=-l}^{l}
    a_{lm}(\omega,t)\,Y_l^m(\theta,\phi),
  \label{eq:eta_SH}
\end{equation}
the terms on the right-hand side (rhs) of Eq.~\eqref{eq:eta_evolution}
can be evaluated using standard angular-momentum algebra.

Since $[\hZ\times\bS]\cdot\nabla_{\!S} = \partial/\partial\phi$, the first term on the rhs gives
\begin{equation}
  \omega[\hZ\times\bS]\cdot\nabla_{\!S}\eta
  = \sum_{l,m} (im\omega)\,a_{lm}\,Y_l^m.
\end{equation}
Because $\PP_0\parallel\hZ$, the second term on the rhs also contributes only a phase:
\begin{equation}
  \mu[\PP_0\times\bS]\cdot\nabla_{\!S}\eta
  = \sum_{l,m} \Bigg[im\mu\big(2\beta-1\big) \frac{\mathcal{I}_1(\kappa)}{\mathcal{I}_0(\kappa)} \Bigg] a_{lm}~Y_l^m.
\end{equation}

We now evaluate the third term on the rhs of Eq.~\eqref{eq:eta_evolution}. Since the order parameter $\PP_\eta$ projects only onto $l=1$ modes, only
$a_{1,\pm1}$ and $a_{1,0}$ enter into the computation. Writing 
$[\PP_\eta\times\bS]\cdot\nabla_{\!S}=i\PP_\eta\cdot\Lhat$,
where \mbox{$\Lhat=-i\bS\times\nabla_{\!S}$}, is the angular momentum operator, and using the ladder operators, 
$\Lhat_\pm Y_l^m=c_{lm}^\pm Y_l^{m\pm1}$ with
$c_{lm}^\pm=\sqrt{l(l+1)-m(m\pm1)}$, the third term couples the $l$-th mode of the perturbation to the $l=1$
components of $\PP_\eta$ acting on $F_0$.
The fourth term is evaluated similarly, with the same operator now acting on $\eta$.
Since $\PP_\eta$ is linear in $\eta$, this contribution is quadratic and generates
terms of the form $a_{1m}\overline{a}_{1m'}$. Crucially, the interaction \emph{only} couples modes with the same $l$, and not with different~$l$'s. Combining these results, we obtain the exact nonlinear evolution equations for the
coefficients of $Y_l^m(\theta,\phi)$ in Eq.~\eqref{eq:eta_SH} (see Appendix~\ref{eq: non lin dyn}). For example, the equation for $l=m=1$ reads as
\begin{widetext}
\begin{equation}
        \frac{\partial a_{1,1}}{\partial t} = -i\bigg[\omega+\mu (2\beta-1)\frac{\mathcal{I}_1(\kappa)}{\mathcal{I}_0(\kappa)}\bigg]a_{1,1} +i\mu \bigg[\beta~g_1(\omega)-(1-\beta)g_2(\omega) \bigg]\frac{\mathcal{I}_1(\kappa)}{\mathcal{I}_0(\kappa)} \overline{a}_{1,1} - i\mu \sqrt{\frac{4\pi}{3}} \bigg[a_{1,1} \overline{a}_{1,0}-a_{1,0}\overline{a}_{1,1}\bigg],\label{eq: a11 evolution}
    \end{equation}
where we have $\overline{a}_{lm} \equiv \int d\omega~a_{lm}$. For general $l,m$, the evolution equation reads
\begin{align}
    \frac{\partial a_{lm}}{\partial t} &= -  \bigg[im\bigg\{\omega+\mu\big(2\beta-1\big) \frac{\mathcal{I}_1(\kappa)}{\mathcal{I}_0(\kappa)}\bigg\}a_{lm} -i \mu \delta_{|m|,1}m \Gamma_l(\omega,\beta,\kappa) \overline{a}_{1,m}\bigg]  \nonumber\\
    &-i\mu \sqrt{\frac{2\pi}{3}}\bigg[\overline{a}_{1,-1} c_{l,m+1}^{-}a_{l,m+1} - \overline{a}_{1,1} c_{l,m-1}^{+}a_{l,m-1}+\sqrt{2}m\overline{a}_{1,0}a_{lm}\bigg].\label{eq: alm evolution}
\end{align}
\end{widetext}

\subsection{Eigenvalue Condition}
\label{sec:eigenvalue}

We focus on the $l=1$ sector, which contains information about the order parameter and therefore about the collective physics.  Writing
$a_{1,1}(\omega,t) = \tilde{a}_{1,1}(\omega)\,e^{-i\Omega t}$, with $\Omega$ being the (complex) frequency of the perturbation,
and retaining only the linear terms in Eq.~\eqref{eq: a11 evolution}, we obtain the dispersion relation determining the quantity $\Omega$ (see Appendix~\ref{app: Critical Point}):
\begin{equation}
  \frac{\Ii(\kappa)}{\Io(\kappa)}\int_{-\infty}^{\infty}d\omega\,
  \frac{\mu\bigl[\beta\,g_1(\omega)-(1-\beta)\,g_2(\omega)\bigr]}
       {\omega + \mu\bigl(2\beta-1\bigr)\tfrac{\Ii(\kappa)}{\Io(\kappa)} - \Omega}
  = 1.
  \label{eq:eigenvalue}
\end{equation}
Instability of the state~\eqref{eq:F0_NS} under dynamical evolution  corresponds to $\mathrm{Im}(\Omega) > 0$.

Equation~\eqref{eq:eigenvalue} is the central result of this paper.
It generalizes the stability analysis of collective oscillations from
a single fixed initial condition to \emph{a full family of stationary
states} parameterized by $(\kappa,\beta)$.

\section{Analytical Results}
\label{sec:analytic}

For a given stationary state $(\kappa,\beta)$ and frequency distribution
$g(\omega)$, Eq.~\eqref{eq:eigenvalue} defines $\Omega$ implicitly.
The integral on the left-hand side is a Cauchy-type transform and must
be analytically continued from the upper half $\Omega$-plane.  For
distributions $g(\omega)$ with support on the real line, one finds:
\begin{enumerate}[noitemsep]
  \item If the integral has no solution with $\mathrm{Im}(\Omega)>0$,
        the state is \emph{linearly stable}.
  \item A solution with $\mathrm{Im}(\Omega)>0$ signals an instability
        that will cause the stationary state to evolve toward a different
        configuration.
\end{enumerate}

\subsection{Special Case I: Two-Beam Model}
\label{sec:two_beams}

In the limit $\kappa\to\infty$, we have $\Ii(\kappa)/\Io(\kappa)\to 1$,
and the two populations concentrate into $\delta$-functions at the
north and south poles.  Taking
$g_1(\omega) = \delta(\omega-\omega_0)$ and
$g_2(\omega) = \delta(\omega+\omega_0)$ (monochromatic beams with
equal and opposite precession frequencies), Eq.~\eqref{eq:eigenvalue}
becomes
\begin{equation}
  \mu\biggl[\frac{\beta}{\omega_0 + \mu\left(2\beta-1\right) - \Omega}
           -\frac{1-\beta}{-\omega_0 + \mu\left(2\beta-1\right)-\Omega}
           \biggr] = 1.
  \label{eq:two_beam}
\end{equation}
Solving this quadratic equation yields
\begin{equation}
  \Omega = \mu \bigg(\beta-\frac{1}{2}\bigg)\pm\,\sqrt{\omega_0^2 - \mu\omega_0
            + \frac{\mu^2(2\beta-1)^2}{4}}.
  \label{eq:Omega_two_beam}
\end{equation}
Instability ($\Omega$ having imaginary component, i.e., the argument of the square
root is negative) occurs when
\begin{equation}
  \omega_0^2 - \mu\omega_0 + \frac{\mu^2(2\beta-1)^2}{4} < 0.
  \label{eq:instability_condition}
\end{equation}

\paragraph{Symmetric Case}

For equal neutrino and antineutrino populations, $\beta=1/2$, and
Eq.~\eqref{eq:Omega_two_beam} simplifies to
$\Omega = \pm\sqrt{\omega_0(\omega_0-\mu)}$.
Instability sets in when $\omega_0 < \mu$, i.e.,
\begin{equation}
  \mu > \omega_0.
\end{equation}
This is the standard instability threshold for the two-beam model,
in agreement with the result of Ref.~\cite{Duan:2007mv,Hannestad:2006nj}.

\paragraph{Asymmetric Case}

For general $\beta$, introducing $\alpha = (1-\beta)/\beta$ and
$\bar\mu \equiv \beta\mu=\mu/(1+\alpha)$, the stability boundary in the
$\bar\mu/\omega_0-\alpha$ plane traces a curve that generalizes the
symmetric result and can be computed analytically:
\begin{equation}
  \bar\mu_{\pm} = \frac{2\omega_0(1+\alpha)}{(1-\alpha)^2}
            \pm \frac{4\omega_0\sqrt{\alpha}}{(1-\alpha)^2}= \frac{2\omega_0}{(1\mp\sqrt{\alpha})^2}.
  \label{eq:critical_mu}
\end{equation}

\subsection{Special Case II: Depolarized State}
\label{sec:uniform_stability}

For the fully depolarized state, we have $\kappa=0$, for which
$\lim_{\kappa\to 0}\mathcal{I}_1(\kappa)/\mathcal{I}_0(\kappa)=0$, and hence, Eq.~\eqref{eq:eigenvalue} is inapplicable. We therefore return to Eq.~\eqref{eq: a11 evolution}, which, to linear order, reduces to
\begin{equation}
    \frac{\partial a_{1,1}}{\partial t}=-i\omega a_{1,1}.
\end{equation}
Thus, each mode undergoes independent precession, with no growth or decay, rendering the fully depolarized state linearly stable. Physically, this can be understood by noting that $\vec{\mathbb{P}}=\vec{0}$ for the fully depolarized state. Consequently, Eq.~\eqref{eq:EoM_OP} implies that each $\vec{\mathbf{S}}_j$ evolves independently, without generating any collective mode.

\subsection{Special Case III: Two-Lorentzian}
\label{sec:two_lorentzian}

We now consider the case $g_1(\omega) =(\sigma/\pi)/[(\omega-\omega_0)^2+\sigma^2] $ and $g_2(\omega) =(\sigma/\pi)/[(\omega+\omega_0)^2+\sigma^2] $. In this case, the computation of Eq.~\eqref{eq:eigenvalue}
depends on the poles of the denominator in the integrand, which in turn depend on the sign of $\Im(\Omega)$. Assuming $\Im(\Omega)<0$, computation of Eq.~\eqref{eq:eigenvalue} gives the allowed solution.
\begin{equation}
    \Omega =\mu \bigg(\beta-\frac{1}{2}\bigg)\mathcal{I}+ i\sigma - \frac{1}{2}\sqrt{(\mathcal{I}\mu-2\omega_0)^2-4\beta(1-\beta)\mathcal{I}^2\mu^2},
\end{equation}
with the condition
\begin{equation}
    (\mathcal{I}\mu-2\omega_0)^2-4\beta(1-\beta)\mathcal{I}^2\mu^2<-4\sigma^2, \label{eq: condition}
\end{equation}
where $\mathcal{I} = \mathcal{I}_1(\kappa)/\mathcal{I}_0(\kappa)$. Similarly, assuming $\Im(\Omega)>0$, computation of Eq.~\eqref{eq:eigenvalue} gives the allowed solution
\begin{equation}
    \Omega =\mu \bigg(\beta-\frac{1}{2}\bigg)\mathcal{I} -i\sigma + \frac{1}{2}\sqrt{(\mathcal{I}\mu-2\omega_0)^2-4\beta(1-\beta)\mathcal{I}^2\mu^2},
\end{equation}
with the condition given in Eq.~\eqref{eq: condition}. In terms of $\alpha \equiv (1-\beta)/\beta$ and $\bar{\mu} = \beta \mu$ we may rewrite Eq.~\eqref{eq: condition} as $\big(\bar{\mu}-\bar{\mu}_-\big)\big(\bar{\mu}-\bar{\mu}_+\big)<0$ with
\begin{equation}
    \bar{\mu}_\pm = \frac{\mathcal{I}_0(\kappa)}{\mathcal{I}_1(\kappa)}\frac{2\omega_0(1+\alpha)}{(1-\alpha)^2}\left[1\pm\sqrt{1-\frac{\big(1-\alpha)^2}{\big(1+\alpha)^2}\left(1+\frac{\sigma^2}{\omega_0^2}\right)}\right]. \label{eq: mu critical lorentzian}
\end{equation}
Clearly, condition~\eqref{eq: condition} is satisfied when $\bar{\mu}_-<\bar{\mu}<\bar{\mu}_+$. Hence, the critical lines enclosing the bipolar oscillation region are given by $\bar{\mu}_\pm$. Note that by setting $\sigma=0$ in Eq.~\eqref{eq: mu critical lorentzian}, we recover the critical points for the two-beam model given in Eq.~\eqref{eq:critical_mu}. Moreover, the critical points $\bar \mu_\pm$ must be real-valued. This imposes an additional constraint on Eq.~\eqref{eq: mu critical lorentzian}: the discriminant must be nonnegative. Requiring the argument of the square root to be positive yields
\begin{equation}
    \alpha>\alpha_{\rm crit}=\frac{\sqrt{\omega_0^2+\sigma^2}-\omega_0}{\sqrt{\omega_0^2+\sigma^2}+\omega_0}.
    \label{eq:alpha_crit}
\end{equation}
\paragraph*{No Bipolar Window}: For $\alpha<\alpha_{\rm crit}$, the quantities $\bar\mu_\pm$ become imaginary, and therefore there is no interval of real $\bar\mu$ satisfying
the instability condition in Eq.~\eqref{eq: condition}. Thus, the
bipolar instability is absent for \emph{all} interaction strengths in this
region. At $\alpha=\alpha_{\rm crit}$, the two boundaries merge, and
the unstable interval disappears. We will refer to
$\alpha<\alpha_{\rm crit}$ as the \emph{no-bipolar-window}, shaded by a gray region in subsequent plots.

\subsection{Connections to Crossing Criteria}
\label{sec:crossing_criterion}

The existence of an unstable solution with \mbox{$\mathrm{Im}(\Omega)>0$} is guaranteed by a necessary and sufficient condition~\cite{Dasgupta:2025quc}. These are analogous to the Penrose criteria~\cite{Penrose1960} for instabilities of plasma.  To state it in our setting, we define the effective spectrum as,
\begin{equation}
  h(\omega)\equiv
  \bar\mu\,\frac{\Ii(\kappa)}{\Io(\kappa)}
  \Big[g_1(\omega) - \alpha\,g_2(\omega)\Big],
  \label{eq:h(omega)}
\end{equation}
and a shifted eigenvalue, 
\begin{equation}
   z\equiv\Omega-\bar\mu\left(1-\alpha\right)\frac{\Ii(\kappa)}{\Io(\kappa)}.
\end{equation}
With these definitions Eq.~\eqref{eq:eigenvalue} takes the form,
\begin{equation}
  \mathcal{D}(z)\equiv
  1+\int_{-\infty}^{\infty} d\omega\;\frac{h(\omega)}{z-\omega}=0.
  \label{eq:crossing_form_D}
\end{equation}
In our setting, the crossing conditions are
\begin{align}
\text{\textit{cond. 1}:}\quad 
& h(\omega_c)=0,\; {\rm with}\; h'(\omega_c)>0,
\\[1pt]
\text{\textit{cond. 2}:}\quad & I_{\rm PV}={\rm P.V.}\int d\omega\,
\frac{h(\omega)}{\omega_c-\omega}<-1,
\label{eq:P2_condition}
\end{align}
where $\omega_c$ is a crossing point of $h(\omega)$, $ h'(\omega_c)$ is the slope at the crossing, and the integral is performed in the principal value (P.V.) sense. These are generalized conditions proposed in Ref.~\cite{Dasgupta:2025quc} for an ultra-relativistic neutrino plasma with a distribution over momentum space \mbox{$(\omega,\bf v)\in\Gamma$}. There, the dispersion relation admits a complex root in the upper half-plane if the contour of the dispersion relation ${\cal D}$ winds around the origin in an anticlockwise fashion, given that these conditions are satisfied.

For the north-south oriented Lorentzian distributions in Eq.~\eqref{eq:F0_NS}, the $g_1(\omega)$ and $g_2(\omega)$ are centered around $\pm\omega_0$ with equal width $\sigma$, the effective spectrum becomes,
\begin{equation}
  h(\omega)\propto g_1(\omega) - \alpha\,g_2(\omega)\,.
  \label{eq:h_eff_Lorent}
\end{equation}
Condition~$\bf\textit{1}$ requires $h(\omega)$ to change sign at some real $\omega_c$.  Setting $h(\omega_c)=0$ gives $g_1(\omega_c)=\alpha\,g_2(\omega_c)$,
which for equal-width Lorentzians reduces to a quadratic in $\omega_c$. A real solution exists if and only if the discriminant is non-negative, which gives the identical condition $\alpha>\alpha_{\rm crit}$ earlier derived from eigenvalue analysis in Eq.~\eqref{eq:alpha_crit}. For $\alpha < \alpha_{\rm crit}$, the effective spectrum $h(\omega)$ has
no crossing, so condition~$\bf\textit{1}$ fails, and by the crossing criterion the stationary state is linearly stable for all values of $\bar\mu$. This is a region where no spectral crossing develops, and no collective instability can exist at any neutrino density. For $\alpha>\alpha_{\rm crit}$, $h(\omega)$ has two real crossings at $\omega_c^{1,2}$ with negative and positive slopes respectively. The corresponding P.V. integrals become
\begin{equation}
    I_{\rm PV}\left(\omega_c^{1}\right)=-\frac{\bar\mu}{\bar\mu_+},\quad
    I_{\rm PV}\left(\omega_c^{2}\right)=-\frac{\bar\mu}{\bar\mu_-},    
\end{equation}
Therefore, within the bipolar window $\bar\mu_-<\bar\mu<\bar\mu_+$ we have
\begin{equation}
    I_{\rm PV}\left(\omega_c^{1}\right)>-1,
    \quad
    I_{\rm PV}\left(\omega_c^{2}\right)<-1.
\end{equation}

\begin{figure*}
    \centering
    \includegraphics[width=0.34\linewidth]{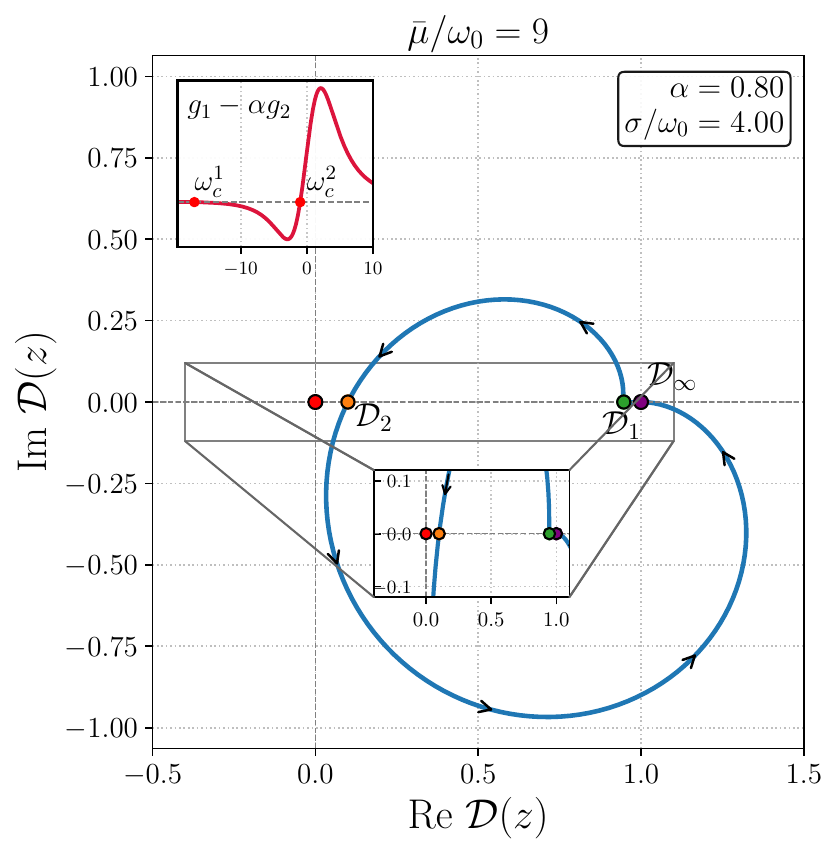}
    \includegraphics[width=0.315\linewidth]{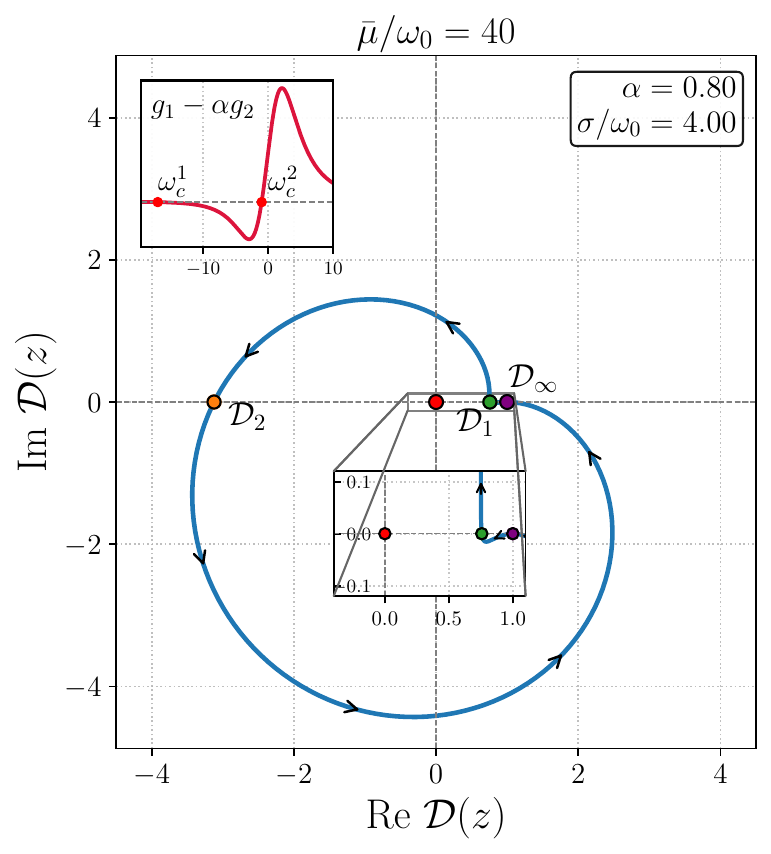}
    \includegraphics[width=0.303\linewidth]{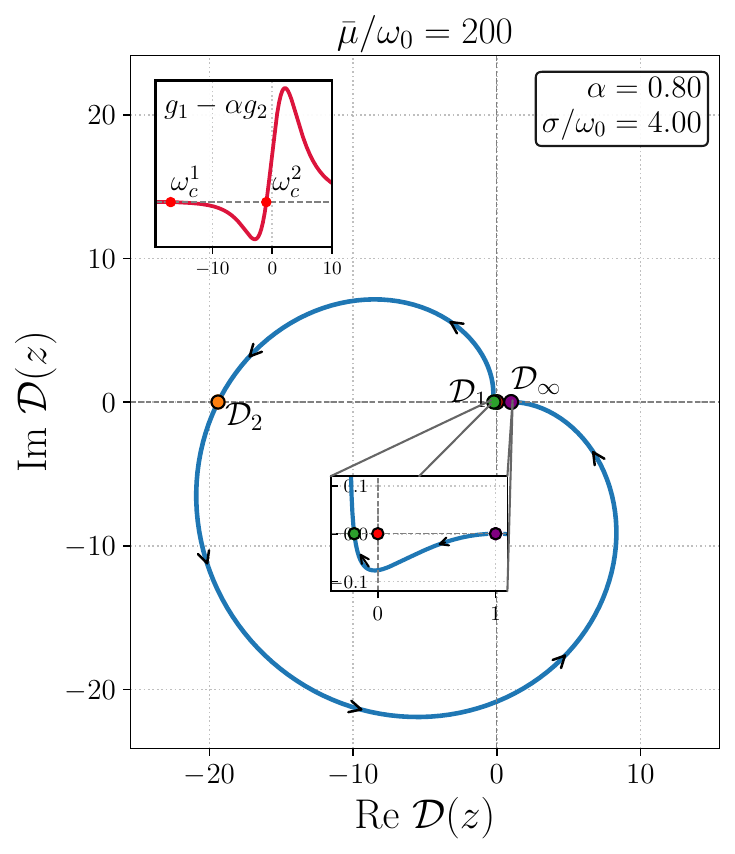}
    \caption{Contours of the dispersion relation ${\cal D}(z)$ for the crossing criterion for the effective two-Lorentzian spectrum $g_1(\omega)-\alpha g_2(\omega)$ shown in the top-left inset. The panels show the contours below, inside, and above the unstable interval. The center insets show a zoomed-in region near the origin, with the contours not encircling, encircling, and not encircling the origin, respectively.  }
    \label{fig:DR_contour_plots}
\end{figure*}

The two intersections of ${\cal D}$ with the real axis lie on opposite sides of the origin, satisfying the instability condition for a spectrum with multiple crossings (see cond. 2b of Ref.~\cite{Dasgupta:2025quc}), and hence predict an unstable mode. For values of $\bar\mu$ outside this region, condition $\bf\textit{2}$ is not satisfied, and both intersections of ${\cal D}$ lie on the same side of the origin, resulting in no instability. Thus, the collective instability is restricted to the interval $\bar\mu_-<\bar\mu<\bar\mu_+$. Fig.~\ref{fig:DR_contour_plots} shows the contour of ${\cal D}(z)$ for $\alpha>\alpha_{\rm crit}$ at several values of $\bar\mu$, spanning the two stable and unstable regions. As expected, the contour encircles the origin only within the unstable bipolar window, while it misses encirclement in the stable (vacuum and synchronized) regions, as shown in the zoomed-in insets.

\section{Numerical Examples}
\label{sec:numerical}

We now illustrate the stability condition derived from Eq.~\eqref{eq:eigenvalue} for
several choices of $g_1(\omega)$ and $g_2(\omega)$ using numerical examples. We measure all quantities in units of $\omega_0$ and time in units of $\omega_0^{-1}$. We use the population-asymmetry parameter $\alpha$, so that $\alpha=1$ corresponds to equal north and south populations, while $\alpha<1$ corresponds to an excess of the north-oriented population. 

\subsection{Two Thick Beams Model} 
\label{sec:phase_diagram} 

We first consider the monochromatic limit, with finite spread around the delta functions centered at $\pm\omega_0$ as, 
\begin{equation}
g_1(\omega)=\delta(\omega-\omega_0), \quad g_2(\omega)=\delta(\omega+\omega_0).
\end{equation}

\begin{figure*}
    \centering
    \includegraphics[width=0.99\linewidth]{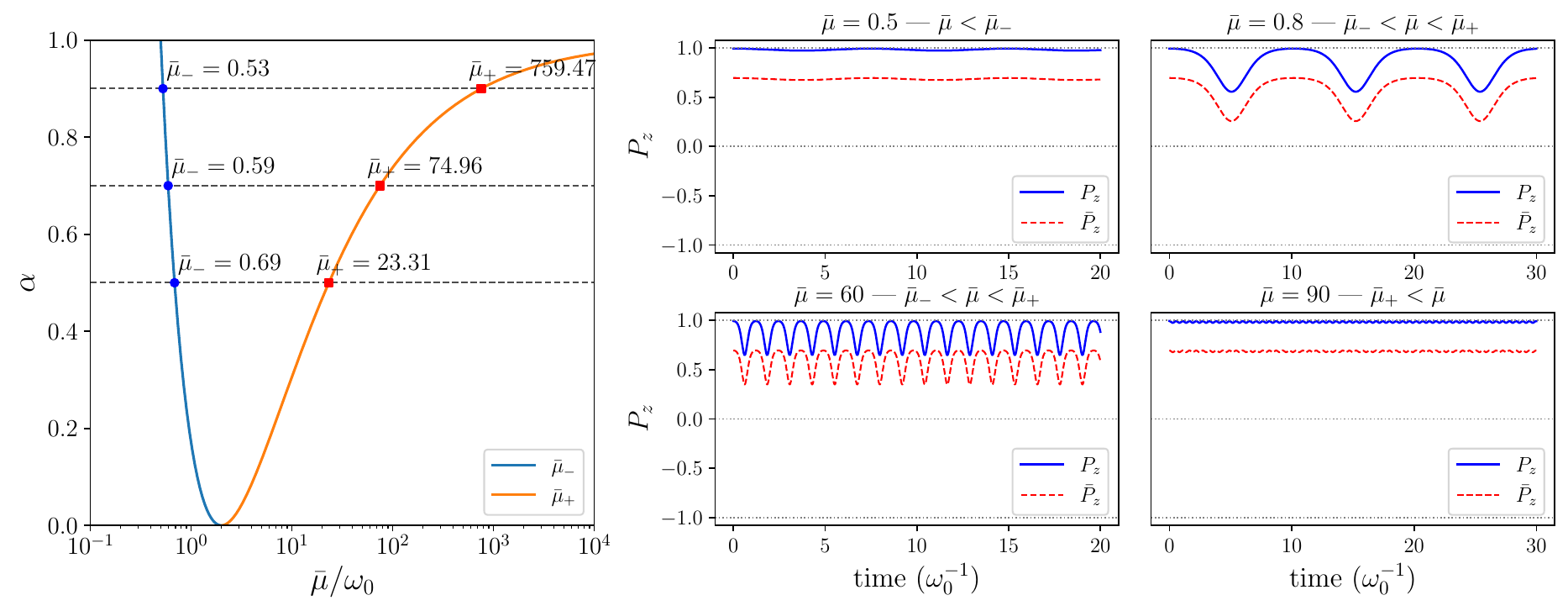}
    \caption{Phase diagram and time evolution for the two thick-beam model. Left panel: The analytically predicted phase diagram in the $\bar\mu/\omega_0-\alpha$ plane, showing the stable and unstable bipolar regions separated by the boundaries $\bar\mu_\pm$ given in Eq.~\eqref{eq:critical_mu}. The horizontal dashed lines mark three representative asymmetry values, $\alpha=0.9,0.7,0.5$, with the corresponding critical points indicated. Right panels: Time evolution of $P_z$ and $\bar P_z$ for $\alpha=0.7$ as the neutrino self-interaction strength $\bar\mu$ is varied. For $\bar\mu<\bar\mu_-$ and $\bar\mu>\bar\mu_+$, the $P_z$ and $\bar P_z$ remain close to their initial values. Within $\bar\mu_-<\bar\mu<\bar\mu_+$, they develop bipolar oscillations, with small dips near the lower boundary and large-amplitude motion deep inside the unstable interval.}
    \label{fig:phase_bipolar_combined}
\end{figure*}

In this case, the eigenvalue equation can be solved analytically, giving the instability condition in Eq.~\eqref{eq:instability_condition} and the boundaries $\bar\mu_-$ and $\bar\mu_+$ in Eq.~\eqref{eq:critical_mu}. For the numerical evolution, we represent the two populations by finite ensembles of unit polarization vectors $\bS$. The north and south cohorts contain $N_+$ and $N_-$ vectors, respectively, with 
\begin{equation} 
\alpha=\frac{N_-}{N_+}, 
\end{equation}
and use $N_+=10^4$ vectors. The vectors $\{\bS^+,\bS^-\}$ are sampled from \begin{equation} 
p_\pm(\cos\theta)=\frac{\kappa}{2\sinh\kappa}\exp(\pm\kappa \cos\theta), 
\end{equation} 
with uniformly sampled azimuthal angles. We take \mbox{$\kappa=1000$}, so each cohort has a small spread around each pole. We then tilt the north (south) cohort away from $+\hZ$ ($-\hZ$), giving a common tilt $\delta=2\theta_{\rm mix}\simeq0.1\,{\rm rad}$. This provides a common transverse seed for the instability. The north cohort is assigned $\omega_j=+\omega_0$, while the south cohort is assigned $\omega_j=-\omega_0$. Each spin is evolved according to Eq.~\eqref{eq:EoM_OP}, with the coupling entering the EoMs given by
$\mu=(1+\alpha)\bar\mu$. The cohort-averaged observables are 
\begin{equation} 
\PP=\frac{1}{N_+}\sum_{j}\bS^+_j, \qquad \bar{ \PP} =\frac{-\alpha}{N_-}\sum_{j}\bS^-_j, 
\end{equation} 
and we plot their longitudinal components $P_z$ and $\bar P_z$. 
For the representative case $\alpha=0.7$, the analytical boundaries are $\bar\mu_-\simeq0.59$ and $\bar\mu_+\simeq75$. In Fig.~\ref{fig:phase_bipolar_combined}, we choose values of $\bar\mu$ below, inside, and above this interval. Within $\bar\mu_-<\bar\mu<\bar\mu_+$, $P_z$ and $\bar P_z$ develop
large-amplitude bipolar oscillations, confirming the instability
predicted by the eigenvalue analysis. Outside this interval, $P_z$ and $\bar P_z$ remain close to their initial values, indicating that both regions are linearly stable.

\subsection{Two-Lorentzian Model}
\label{sec:smooth_g}
We next consider a smooth frequency distribution by replacing the monochromatic beams with two-Lorentzian distributions,
\begin{align}
g_1(\omega)&=\frac{\sigma/\pi}{(\omega-\omega_0)^2+\sigma^2},\nonumber\\
g_2(\omega)&=\frac{\sigma/\pi}{(\omega+\omega_0)^2+\sigma^2}.
\end{align}
The width $\sigma$ introduces a spread of vacuum frequencies around $\pm\omega_0$. The north and south polarization vectors are sampled from the same angular distributions as above, again with the same tilt angle and $\kappa=1000$. We use $N_+=10^4$ vectors in the north cohort and $N_-=\alpha N_+$ vectors in the south cohort.

The frequencies of the north and south cohorts are sampled independently from Lorentzian distributions centered at $+\omega_0$ and $-\omega_0$, respectively. For numerical stability, the long Lorentzian tails are truncated at 50 widths around each peak, $|\omega\mp\omega_0|\leq50\sigma$
for the two cohorts, respectively.

\begin{figure*}
    \centering
    \includegraphics[width=0.99\linewidth]{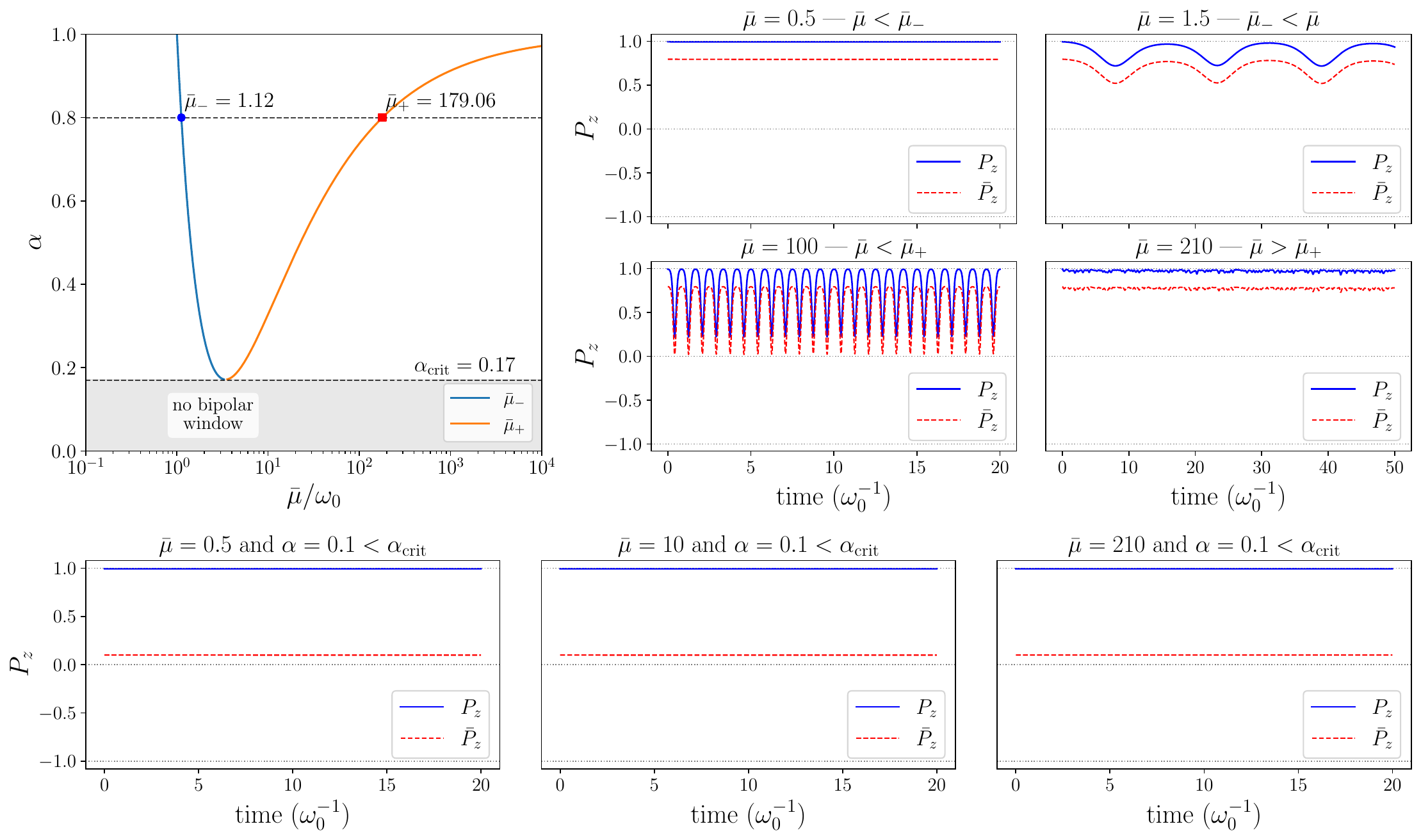}
    \caption{Phase diagram and time evolution for the two-Lorentzian model. The top-left panel shows the analytical instability boundaries in the $\bar\mu/\omega_0-\alpha$ plane, obtained for the two-Lorentzian spectrum given in Eq.~\eqref{eq: mu critical lorentzian}. The upper-right panels show the time evolution of $P_z$ and $\bar P_z$ for $\alpha=0.8$, and $\bar\mu$ is varied across the stable and unstable regimes. The lower panels show the evolution for $\alpha=0.1$ in the no-bipolar-window (shaded in gray); no instability appears even as $\bar\mu$ is varied across a large range.}
    \label{fig:finite_lorentzian_phase_panels}
\end{figure*}

In Fig.~\ref{fig:finite_lorentzian_phase_panels}, we show an example with $\sigma/\omega_0=1$. For \mbox{ $\alpha=0.8$}, the critical points are real, and the behavior of $P_z$ and $\bar P_z$ is qualitatively the same as in the two thick-beam case. The longitudinal components remain close to their initial values for
$\bar\mu<\bar\mu_-$ and $\bar\mu>\bar\mu_+$, and develop bipolar oscillations
within $\bar\mu_-<\bar\mu<\bar\mu_+$. Thus, these observables verify the linearly unstable interval. For $\sigma/\omega_0=1$, Eq.~\eqref{eq:alpha_crit} gives
$\alpha_{\rm crit}\simeq0.17$. Below this value, the two instability boundaries are no longer real, and the bipolar window does not exist (no-bipolar-window). For $\alpha=0.1$, the lower panels do not show bipolar oscillations over the explored range of~$\bar\mu$.

\begin{figure*}
    \centering
    \includegraphics[width=1\linewidth]{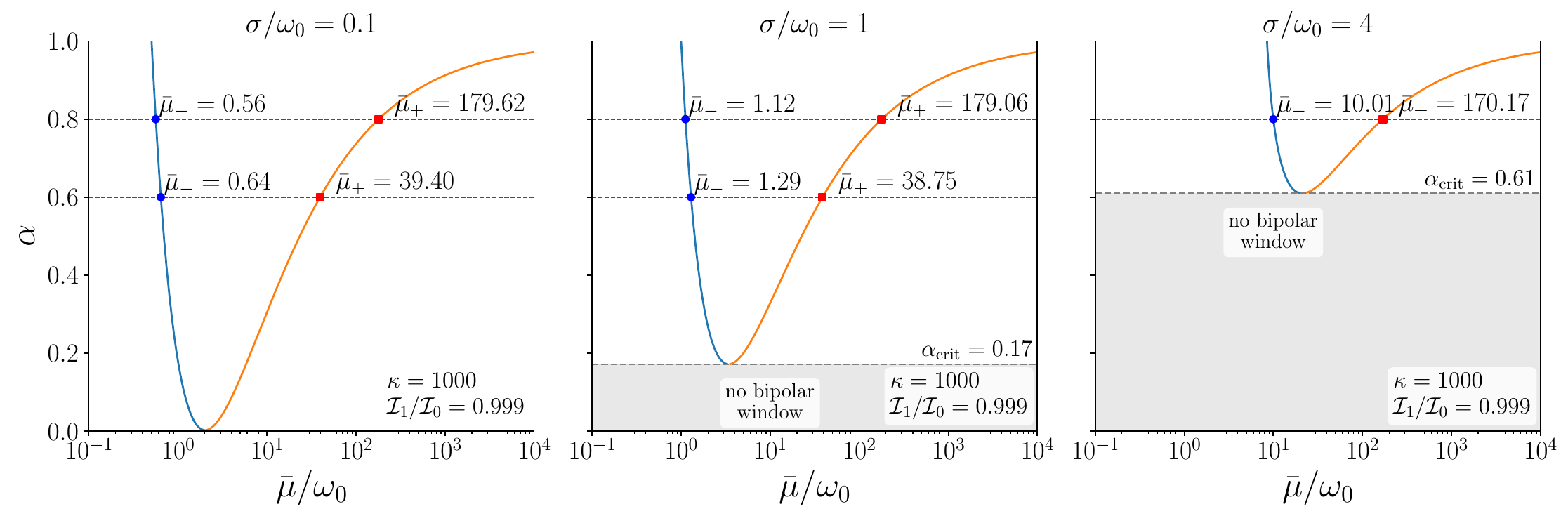}
    \caption{Phase diagrams for the north-south cohort with two-Lorentzian distributions centered at $\pm\omega_0$. Here $\kappa$ and $\Ii(\kappa)/\Io(\kappa)$ are fixed for all the panels, while the Lorentzian width is varied across the panels: $\sigma/\omega_0=0.1, 1, 4$. As the width of the frequency distribution increases, the bipolar window narrows. For sufficiently broad widths, the no-bipolar-window becomes significantly large in the phase space, and no bipolar instability is predicted.}
    \label{fig:phase_diagram_kappa_sigma}
\end{figure*}

In Fig.~\ref{fig:phase_diagram_kappa_sigma}, for small width $\sigma/\omega_0=0.1$, the phase diagram is similar to the two thick-beam case. Increasing the width $\sigma$ shifts the lower boundary $\bar\mu_-$ to larger values, while the upper boundary $\bar\mu_+$ shifts less at large $\alpha$. The interval of $\bar\mu$ over which bipolar motion is expected becomes narrower. For sufficiently broad spectra, the boundaries $\bar\mu_\pm$ are no longer
real below $\alpha_{\rm crit}$, and no bipolar instability window exists. For example, when $\sigma/\omega_0=4$, we find \mbox{$\alpha_{\rm crit}\simeq0.61$}, which has a large region without a bipolar window. This follows from the crossing criterion of Sec.~\ref{sec:crossing_criterion}: $h(\omega)$ develops real crossings only for $\alpha>\alpha_{\rm crit}$. For $\alpha<\alpha_{\rm crit}$, the crossing disappears, and no bipolar instability occurs.

\begin{figure*}
    \centering
    \includegraphics[width=0.6\linewidth]{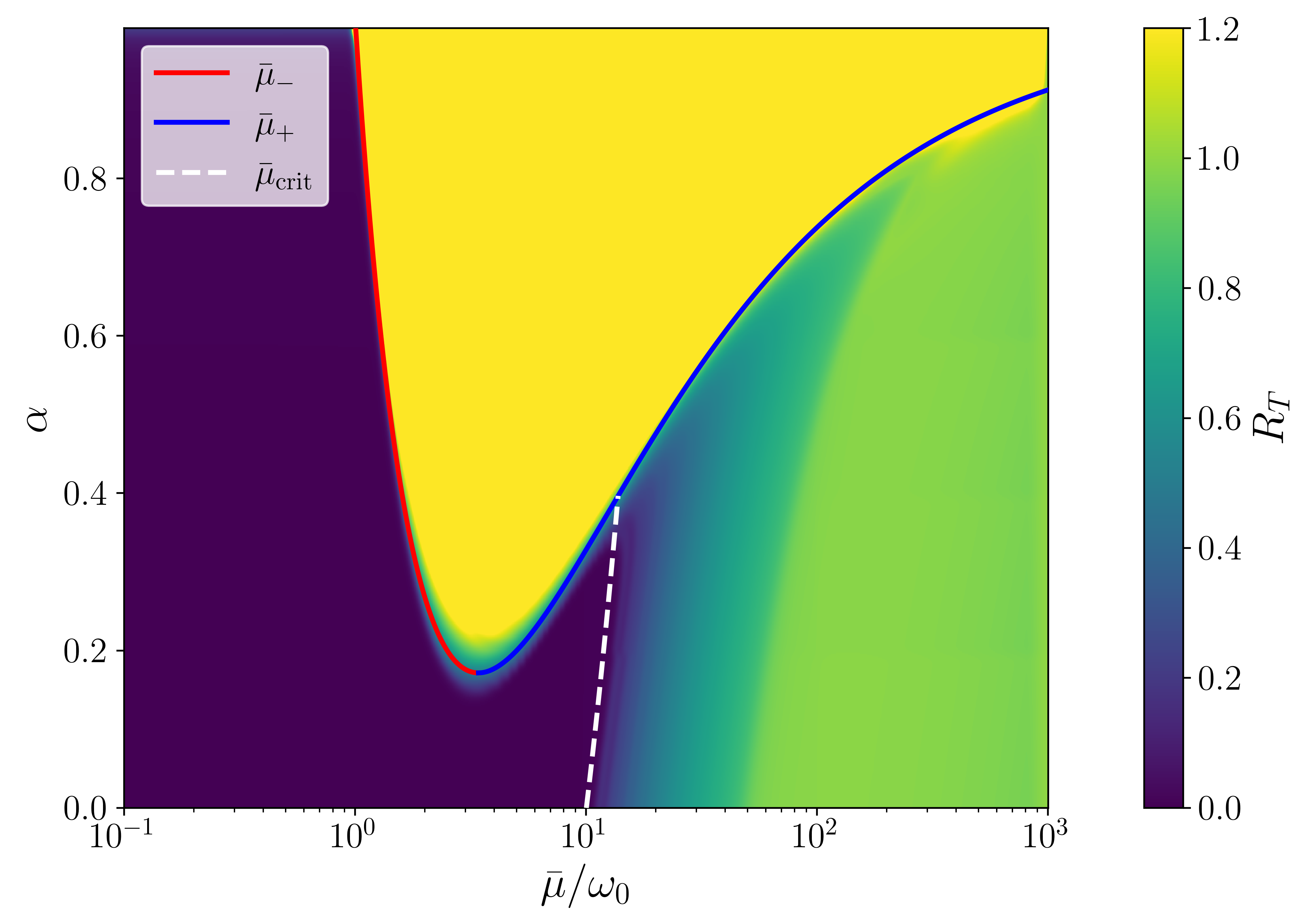}
    \caption{Combined phase diagram in the
    $\bar\mu/\omega_0-\alpha$ plane for the two-Lorentzian spectrum. The red and blue solid curves denote the analytical instability boundaries $\bar\mu_-$ and $\bar\mu_+$, respectively, while the white dashed curve shows the synchronization threshold $\bar\mu_{\rm crit}$, continued into the instability region for numerical comparison. Dark regions correspond to $R_T\simeq0$, while nonzero values indicate collective transverse motion. The color scale is clipped at $R_T=1.2$ for visual clarity. In the simulations we use $\omega_0=1$, $\sigma/\omega_0=1$, $\delta=2\theta_{\rm mix}=0.1$ rad, and $\kappa=1000$, with $1.5\times10^4$ frequency modes per cohort and a total of $4\times10^5$ sampled points in the $\bar\mu/\omega_0-\alpha$ plane.}
    \label{fig:RT_density_plot}
\end{figure*}


\subsection{Transverse Dephasing and Synchronization}
\label{sec:transverse_dephasing}
The longitudinal components $P_z$ and $\bar P_z$ diagnose bipolar motion through large-amplitude oscillations, but remain nearly stationary in both stable regimes. They therefore do not distinguish a transversely
dephased, i.e., vacuum-like state at low interaction strength, from synchronized motion at large interaction strength. To separate these behaviors, we consider the late-time transverse polarization through the order parameter
\begin{equation}
    R_T=
    \left\langle
    \frac{|P_\perp(t)|}{|P_\perp(0)|}
    \right\rangle_{\rm late},
    \label{eq:RT_definition}
\end{equation}
where $|P_\perp|=\sqrt{P_x^2+P_y^2}$ is the transverse component of the total polarization vector $\PP_{\rm tot}$, and $\langle\cdot\rangle_{\rm late}$ refers to an average over a few cycles at late times of the evolution. 
For the initial state, the resultant transverse component of the north-south cohorts is
\begin{equation}
    P_\perp(0)=Q(\kappa)\frac{1-\alpha}{1+\alpha}\sin\delta,
    \label{eq:Pperp_initial}
\end{equation}
where $Q(\kappa)={\Ii(\kappa)}/{\Io(\kappa)},$ and $\delta=2\theta_{\rm mix}$ is the common initial tilt. In the single-cohort and perfectly localized limit, $\alpha=0$ and $Q(\kappa)\to1$, the instantaneous ratio entering Eq.~\eqref{eq:RT_definition} reduces to
$R(t)=P_\perp(t)/\sin2\theta_{\rm mix}$.

The interpretation of $R_T$ is straightforward. A value $R_T=0$ indicates transverse dephasing, where different frequency modes cancel in the collective sum, while \mbox{$R_T>0$} indicates collective transverse motion. In the presence of spectral crossings, however, $R_T$ alone does not distinguish synchronized motion from bipolar oscillations; rather, it distinguishes the two stable regimes.

For $\alpha<\alpha_{\rm crit}$, where no bipolar instability exists, the onset of nonzero $R_T$ directly indicates synchronization. A critical value of interaction strength can be obtained from the self-consistency conditions for sustained late-time collective precession. For the two-Lorentzian spectrum, one finds
\begin{equation}
    \bar\mu_{\rm crit}=\frac{\sin\delta}{Q(\kappa)\sigma\left(D_+^{-1}-\alpha D_-^{-1}\right)},
    \label{eq:mucrit_main}
\end{equation}
where $D_\pm=(\omega_r\mp\omega_0)^2+\sigma^2$, with $\omega_r$ determined in Appendix~\ref{app:RT_two_lorentzian}. For $\alpha=0$, substituting the corresponding value of $D_+$,
this reduces to
\begin{equation}
    \bar\mu_{\rm crit}=\frac{\sigma}{Q(\kappa)\sin\delta}.
    \label{eq:mucrit_alpha0}
\end{equation}
Thus, decreasing the initial tilt shifts the synchronization scale to larger interaction strength, with
$\bar\mu_{\rm crit}\rightarrow\infty$ as $\delta\rightarrow0$. This agrees with the vanishing mixing angle limit discussed in Ref.~\cite{Raffelt:2010za},
where a spectrum with infinite tails remains completely dephased for any finite interaction strength. Note that the order parameter therein was introduced for the stable regime, whereas here $R_T$ involves a late-time average and is extended into the bipolar instability region.

We compute $R_T$ for crossed and uncrossed spectra, over the $\bar\mu/\omega_0-\alpha$ plane. Fig.~\ref{fig:RT_density_plot} shows the resulting $R_T$ together with the synchronization threshold $\bar\mu_{\rm crit}$ and the instability boundaries $\bar\mu_\pm$. At small $\bar\mu$, $R_T=0$, while at sufficiently large interaction strength the system develops synchronized transverse motion approaching $R_T\simeq1$. Within the bipolar interval, pendular motion gives \mbox{$R_T>1$}. 

For $\alpha<\alpha_{\rm crit}$, the numerical transition from $R_T=0$ to $R_T>0$ follows $\bar\mu_{\rm crit}$. Immediately above $\alpha_{\rm crit}$, however, the end of the bipolar instability does not coincide with the onset of synchronization. Transverse dephasing persists beyond $\bar\mu_+$, producing an intermediate region $\bar\mu_+<\bar\mu<\bar\mu_{\rm crit}$ until $\bar\mu_{\rm crit}$ meets $\bar\mu_+$. Thus, immediately beyond $\bar\mu_+$, a linearly stable system need not yet be synchronized.

There is no a priori reason for $\bar\mu_+$ and the transverse synchronization threshold $\bar\mu_{\rm crit}$ to coincide, since they diagnose different transitions. Indeed, in physical systems, different order parameters may be designed to probe different aspects of the system and therefore need not exhibit critical behavior at the same point. A familiar example is a material that undergoes magnetic ordering at one temperature and becomes superconducting only at a lower temperature. Nevertheless Fig.~\ref{fig:RT_density_plot} shows that the
continued $\bar\mu_{\rm crit}$ line meets the bipolar boundary $\bar\mu_+$ near $\alpha\simeq0.4$, beyond which the bipolar to synchronized transition occurs directly across the upper boundary.

\paragraph*{Relation to Kinematic Decoherence:}
The transverse dephasing discussed here is a form of kinematic decoherence among modes with different vacuum frequencies. In our spatially homogeneous, single-angle system, the individual polarization vectors evolve coherently, but their transverse components dephase in the collective sum, giving $R_T\simeq0$.

Kinematic decoherence arises in slow collective oscillations from multi-angle effects, where polarization vectors associated with different trajectories lose their collective coherence~\cite{Raffelt:2007yz,EstebanPretel:2007ec}. A related loss of macroscopic coherence has been studied in the context of fast flavor conversions~\cite{Capozzi:2019lso,Bhattacharyya:2020dhu,Bhattacharyya:2020jpj,Bhattacharyya:2022eed}. In spatially inhomogeneous and anisotropic systems, nonlinear flavor evolution can lead to fast flavor depolarization through transverse relaxation in flavor space. In particular, the instability can transfer power to higher angular multipoles and toward increasingly fine spatial and angular scales. This redistribution of power across fine-grained modes damps the macroscopic polarization vector, leading to a strongly reduced coarse-grained polarization and decoherence. Several other works have explored aspects of nonlinear relaxation, macroscopic decoherence, and the emergence of quasi-steady configurations in collective flavor evolution~\cite{Martin:2019gxb,Johns:2020qsk,Richers:2021xtf,Wu:2021uvt,Nagakura:2022kic,Goimil-Garcia:2025ozm,Liu:2025muc}.

\section{Discussion}
\label{sec:discussion}

In this paper we have developed a distributional framework for the stability
analysis of collective neutrino oscillations.  The key advance over the
standard linear stability analysis is that our approach is valid for
\emph{any} initial distribution $F_0(\bS,\omega)$, not only for
distributions close to a flavor-coherent state.  Starting from the
Fokker--Planck equation \eqref{eq:FP}, we identified a broad class of
azimuthally symmetric stationary solutions \eqref{eq:F0_NS} parameterized
by the concentration $\kappa$ and the population asymmetry $\alpha$.  The
linearized stability of these states is governed by the eigenvalue
condition \eqref{eq:eigenvalue}, which reduces in the two-beam limit to
the known synchronization thresholds of the collective oscillation
literature and in the incoherent limit to the Kuramoto stability criterion.
 The main results of our analysis may be summarized as follows.

\begin{enumerate}

\item We develop a distributional stability framework for collective neutrino oscillations, in which the dynamics is formulated in terms of a one-body distribution on the flavor sphere. Performing linear stability analyses around the north-south oriented cohort of polarizations yields an eigenvalue condition for the instability growth rate. This method extends stability analysis beyond the usual off-diagonal flavor coherence to general distributions on the Bloch sphere.

\item In the monochromatic two-beam limit, the eigenvalue condition reproduces the known lower and upper instability boundaries $\bar\mu_-$ and $\bar\mu_+$. For the two-Lorentzian spectrum, these boundaries can again be obtained analytically and extend the bipolar instability window to continuous frequency distributions. A finite spectral width introduces a critical asymmetry $\alpha_{\rm crit}$, below which the two boundaries become complex and no bipolar instability exists.

\item The same instability condition admits an equivalent interpretation in terms of spectral crossings and Nyquist curves. For the two-Lorentzian model, the effective distribution $h(\omega)\propto g_1(\omega)-\alpha g_2(\omega)$ develops real crossings precisely for $\alpha>\alpha_{\rm crit}$ while the corresponding P.V. conditions confirm the same interval $\bar\mu_-<\bar\mu<\bar\mu_+$. Thus, the eigenvalue analysis and the crossing criterion provide a consistent description of the instability.

\item Direct numerical evolution confirms these analytical predictions. Within the interval $\bar\mu_-<\bar\mu<\bar\mu_+$, the longitudinal polarizations $P_z$ and $\bar P_z$ develop bipolar motion, whereas outside this interval they remain close to their initial values. This agreement holds in both the spectra considered.

\item Linear stability alone does not distinguish the two stable regimes on either side of the bipolar window. We therefore employ a late-time transverse parameter $R_T$ to distinguish the vacuum-like dephased state, with $R_T\simeq0$, from collective transverse motion, with $R_T>0$. For synchronized motion, $R_T$ saturates to unity at large $\bar\mu$, while inside the bipolar region pendular motion produces large transverse motion, and $R_T$ exceeds unity.

\item The combined longitudinal and transverse diagnostics reveal an additional region of the $\bar\mu/\omega_0-\alpha$ phase space. Immediately above $\alpha_{\rm crit}$, numerical evolution shows that transverse dephasing persists beyond the upper bipolar boundary, giving a region $\bar\mu_+<\bar\mu<\bar\mu_{\rm crit}$. This region disappears when the synchronization threshold $\bar\mu_{\rm crit}$ meets $\bar\mu_+$; beyond this point, the evolution passes directly from bipolar motion to synchronized transverse motion across the upper instability boundary.

\end{enumerate}

Several extensions are worth pursuing.
First, the present analysis is restricted to spatially homogeneous,
isotropic configurations.  Restoring the angular dependence
$(1-\mathbf{v}\cdot\mathbf{v}')$ in the neutrino--neutrino potential and
allowing spatial gradients would be the natural next step toward
realistic supernova conditions. With the notation of
Sec.~\ref{sec:FP}, the multi-angle extension amounts to promoting the
one-body density to $F(\bS,\omega,{\bf v},{\bf x},t)$, with the stability
analysis of Sec.~\ref{sec:stability} generalizing to an eigenvalue
condition carrying an additional angular integral over the
direction-resolved mean field. For a sufficiently narrow angular distribution, $\kappa\gg1$, the factor $1-{\bf v}\cdot{\bf v'}$ varies little between different mode pairs, such that it simply rescales $\mu$, with the spread in the effective coupling remaining small compared with its mean value. Multi-angle effects may introduce additional decoherence. Second, a fully nonlinear analysis -- going beyond linear stability to
understand whether unstable modes saturate, and at what amplitude -- remains
an important open problem. The mechanical analogy developed in
Refs.~\cite{Dasgupta:2017oko, Padilla-Gay:2021haz, Fiorillo:2026lyz} for fast conversions suggests that such analyses may be tractable. Third, collisions may provide a connection between the distributional description and stochastic resetting in statistical systems~\cite{PhysRevLett.106.160601, Evans2020, Bressloff2024, PhysRevE.109.064137, np7q-hxld, d82g-qyvf}. In a coarse-grained picture, collisional interactions may act as ``noise'' that resets and irregularly drives the system away from coherent evolution and toward a decohered flavor distribution~\cite{Capozzi:2018clo, Johns:2021qby, Xiong:2022zqz}. Incorporating this provides a route to study how stochastic relaxation modifies the stability boundaries in realistic supernova conditions.

We hope that the framework presented here provides a useful complement to existing analytical and numerical approaches, and that the mechanical analogy with synchronization physics offers new intuition for the rich phase structure of collective neutrino oscillations.

\section*{Acknowledgements}
This work is supported by the Department of Atomic Energy, Government of India, under Project Identification Number RTI-4012 and RTI-4013. Computations were carried out
on the computing clusters at the Department of Theoretical Physics, TIFR, Mumbai. We thank Kapil Ghadiali and Ajay Salve for computational support.


\appendix

\section{Derivation of Eq.~\texorpdfstring{\eqref{eq:P0}}{Eq.~(P0)}}
\label{app: derivation}
We start with the well-known expansion~\cite{DLMF}
\begin{equation}
    e^{i\kappa\cos{\theta}}=\sum_{l=0}^\infty (2l+1)i^lj_l(\kappa)P_l(\cos{\theta)},
    \label{eq:app1-eq1}
\end{equation}
where $j_l$ are the spherical Bessel functions of the first kind and $P_l$ are the Legendre polynomials. Now, by definition, the spherical harmonics $Y_l^m(\theta,\phi)$ for $m=0$ satisfy~\cite{DLMF}
\begin{equation}
    Y_l^0(\theta,\phi) = \sqrt{\frac{2l+1}{4\pi}}P_l(\cos{\theta)}.
\end{equation}
Using this as well as replacing $\kappa$ by $-i\kappa$ in Eq.~\eqref{eq:app1-eq1}, we obtain
\begin{equation}
    e^{\kappa\cos{\theta}}=\sum_{l=0}^\infty \sqrt{4\pi(2l+1)}\mathcal{I}_l(\kappa)Y_l^0,
\end{equation}
where we have the modified spherical Bessel function of the first kind, $\mathcal{I}_l(\kappa) = i^lj_l(-i\kappa)$. Similarly, we have
\begin{align}
    e^{-\kappa\cos{\theta}}&=\sum_{l=0}^\infty \sqrt{4\pi(2l+1)}\mathcal{I}_l(-\kappa)Y_l^0 \nonumber\\
    &=\sum_{l=0}^\infty(-1)^l \sqrt{4\pi(2l+1)}\mathcal{I}_l(\kappa)Y_l^0,
\end{align}
where we have used $\mathcal{I}_l(-\kappa)=(-1)^l\mathcal{I}_l(\kappa)$.
Using the above results, we obtain from Eq.~\eqref{eq:F0_NS} that
\begin{align}
F_0\left(\vec{\mathbf{S}},\omega\right) 
    &= \frac{1}{2\pi} \frac{\kappa}{ \sinh{\kappa}}\sum_{l=0}^\infty \sqrt{\pi(2l+1)}\Big[\beta g_1(\omega) \nonumber\\
    &+ (1-\beta)(-1)^lg_2(\omega)\Big]\mathcal{I}_l(\kappa)Y_l^0.
    \label{eq:F0-expansion-app}
\end{align}

Let us now compute the order parameter for the above state. We have from Eq.~\eqref{eq:OP_F} that
\begin{align}
    \vec{\mathbb{P}}_0\equiv \vec{\mathbb{P}}_{F=F_0}&= \int d\omega'\int \vec{\mathbf{S}}'~F_0\left(\vec{\mathbf{S}}',\omega'\right)~d\Omega'. \label{eq: P F0}
\end{align}
From the definition, $\vec{\mathbf{S}}$ may be expressed in terms of the spherical harmonics as
\begin{equation}
    \vec{\mathbf{S}} = 
    \sqrt{\frac{2\pi}{3}}
    \begin{bmatrix}
        Y_1^{-1}-Y_1^1\\
        iY_1^{-1}+iY_1^1\\
        \sqrt{2}Y_1^0 \label{eq: S in Ylm}
    \end{bmatrix}.
\end{equation}
Using the orthonormality condition of the spherical harmonics, $\int d\Omega\,
Y_\ell^{m*}(\theta,\phi)\,
Y_{\ell'}^{m'}(\theta,\phi)
=
\delta_{\ell\ell'}\delta_{mm'}$, with * denoting complex conjugation, and performing the integrals in Eq.~\eqref{eq: P F0}, we obtain 
\begin{align}
    \vec{\mathbb{P}}_0  = \bigg[2\beta-1\bigg] \frac{\mathcal{I}_1(\kappa)}{\mathcal{I}_0(\kappa)} \hat{\mathbf{z}},
    \label{eq:app1-P0}
\end{align}
which is Eq.~\eqref{eq:P0} of the main text.

\section{Proof of Stationarity of the State~\eqref{eq:F0_NS}}
\label{app: proof stationarity}
We show that $F_0\left(\vec{\mathbf{S}},\omega\right)$ in Eq.~\eqref{eq:F0_NS} is a stationary solution of the continuity equation~\eqref{eq:FP}.
In terms of spherical polar coordinates, we have
\begin{equation}
    \vec{\nabla}_\mathcal{S} \equiv \hat{\theta}\frac{\partial}{\partial \theta} + \hat{\phi} \frac{1}{\sin{\theta}}\frac{\partial }{\partial \phi},
\end{equation}
which gives
\begin{equation}
    \vec{\nabla}_\mathcal{S}F_0\left(\vec{\mathbf{S}},\omega\right) = \hat{\theta}\frac{\partial F_0}{\partial \theta}.
\end{equation}
Using this, we obtain the second term in the continuity equation~\eqref{eq:FP} as 
    \begin{align}
    \Big[\hat{\mathbf{z}} \times \vec{\mathbf{S}}\Big]\cdot \vec{\nabla}_\mathcal{S}F_0 = \sin{\theta}~\hat{\phi} \cdot \vec{\nabla}_\mathcal{S}F_0 = 0,
\end{align}
since $\vec{\nabla}_\mathcal{S}F_0$ is along the $\hat{\theta}$ direction. Similarly, the other term in the continuity equation gives
\begin{equation}
    \Big[\vec{\mathbb{P}} \times \vec{\mathbf{S}}\Big]\cdot \vec{\nabla}_\mathcal{S}F_0 =  \bigg[2\beta-1\bigg] \frac{\mathcal{I}_1(\kappa)}{\mathcal{I}_0(\kappa)}\Big[\hat{\mathbf{z}} \times \vec{\mathbf{S}}\Big]\cdot \vec{\nabla}_\mathcal{S}F_0=0.
\end{equation}
Here, we have used Eq.~\eqref{eq:app1-P0}. Thus, $F_0\left(\vec{\mathbf{S}},\omega\right)$ in Eq.~\eqref{eq:F0_NS} is a stationary solution of the continuity equation~\eqref{eq:FP}. 

\section{Derivation of Eq.~\eqref{eq: a11 evolution}}
\label{eq: non lin dyn}

We start with the expansion~\eqref{eq:eta_SH}, and evaluate 
 each of the terms in Eq.~\eqref{eq:eta_evolution}. Using
\begin{equation}
    \left[\hat{\mathbf{z}}\times\vec{\mathbf{S}} \right]\cdot\vec{\nabla}_\mathcal{S} = \frac{\partial}{\partial \phi},
\end{equation}
the first term on the rhs gives
\begin{align}
    \omega \Big[~\hat{\mathbf{z}} \times \vec{\mathbf{S}}\Big]\cdot \vec{\nabla}_\mathcal{S}\eta=\sum_{l=1}^\infty\sum_{m=-l}^l \Big[im\omega\Big] a_{lm}~Y_l^m.
\end{align}
Similarly, the second term gives
\begin{align}
    &{\mu}\Big[~\vec{\mathbb{P}}_0 \times \vec{\mathbf{S}}\Big]\cdot \vec{\nabla}_\mathcal{S}\eta \nonumber \\
    &= {\mu}\big(2\beta-1\big) \frac{\mathcal{I}_1(\kappa)}{\mathcal{I}_0(\kappa)} \Big[\hat{\mathbf{z}} \times \vec{\mathbf{S}}\Big]\cdot \vec{\nabla}_\mathcal{S}\eta\nonumber\\
    &= \sum_{l=1}^\infty\sum_{m=-l}^l \Bigg[im\mu\big(2\beta-1\big) \frac{\mathcal{I}_1(\kappa)}{\mathcal{I}_0(\kappa)} \Bigg] a_{lm}~Y_l^m.
\end{align}

We now focus on the third term on the rhs, namely, the term $\mu\bigl[\PP_\eta\times\bS\bigr]\cdot\nabla_{\!S}F_0$. To this end, we first compute the order parameter $\vec{\mathbb{P}}_\eta$ of the perturbation. Substituting the expansion of $\eta$ from Eq.~\eqref{eq:eta_SH} into the definition of $\vec{\mathbb{P}}_\eta$ given after Eq.~\eqref{eq:eta_evolution}, we obtain
\begin{align}
    \vec{\mathbb{P}}_\eta &=\int d\omega'\int \vec{\mathbf{S}}'~\eta\left(\vec{\mathbf{S}}',\omega',t\right)~d\Omega'\nonumber\\
    &=\sqrt{\frac{2\pi}{3}}\sum_{l=1}^\infty\sum_{m=-l}^l \int d\omega'a_{lm}(\omega',t) \nonumber\\
    &\times\begin{bmatrix}
        \int d\Omega'Y_1^{-1}Y_l^m-\int d\Omega'Y_1^1Y_l^m\\
        i\int d\Omega'Y_1^{-1}Y_l^m+i\int d\Omega'Y_1^1Y_l^m\\
        \sqrt{2}\int d\Omega'Y_1^0Y_l^m
    \end{bmatrix},
\end{align}
where we have used the expression of $\vec{\mathbf{S}}$ in terms of spherical harmonics from Eq.~\eqref{eq: S in Ylm}. Using the orthonormality condition of spherical harmonics, we simplify the above equation into
\begin{align}
    \vec{\mathbb{P}}_\eta =\sqrt{\frac{2\pi}{3}}
    \begin{bmatrix}
        \overline{a}_{1,-1}-\overline{a}_{1,1}\\
        -i\overline{a}_{1,-1}-i\overline{a}_{1,1}\\
        \sqrt{2}\overline{a}_{1,0}
    \end{bmatrix},
    \label{eq:Peta-app}
\end{align}
where for compactness, we use the notation $\overline{a}_{lm} =\int d\omega'a_{lm}(\omega',t)  $. Now, we have the definition of the angular momentum operator
\begin{eqnarray}
    \widehat{\mathbf{L}} = -i\vec{\mathbf{S}}\times\vec{\nabla}_\mathcal{S},
\end{eqnarray}
so that the third term may be written as
\begin{equation}
   \Big[\vec{\mathbb{P}}_\eta \times\vec{\mathbf{S}}
    \Big]\cdot \vec{\nabla}_\mathcal{S}
    = \vec{\mathbb{P}}_\eta \cdot \Big[\vec{\mathbf{S}}\times\vec{\nabla}_\mathcal{S}\Big] = i\vec{\mathbb{P}}_\eta \cdot  \widehat{\mathbf{L}}.
\end{equation}
Evaluating with the use of Eq.~\eqref{eq:Peta-app}, we have
\begin{align}
    &\Big[\vec{\mathbb{P}}_\eta \times\vec{\mathbf{S}}
    \Big]\cdot \vec{\nabla}_\mathcal{S}\nonumber \\
    &=i\vec{\mathbb{P}}_\eta \cdot  \widehat{\mathbf{L}}= iP_x\widehat{L}_x+iP_y\widehat{L}_y+iP_z\widehat{L}_z\nonumber\\
    &= i \sqrt{\frac{2\pi}{3}} \Big[\overline{a}_{1,-1}-\overline{a}_{1,1}\Big]\widehat{L}_x+ \sqrt{\frac{2\pi}{3}} \Big[\overline{a}_{1,-1}+\overline{a}_{1,1}\Big]\widehat{L}_y\nonumber\\
    &+i \sqrt{\frac{4\pi}{3}}\overline{a}_{1,0}\widehat{L}_z\nonumber\\
    &=i\sqrt{\frac{2\pi}{3}} \overline{a}_{1,-1} \Big(\widehat{L}_x-i\widehat{L}_y\Big)-i\sqrt{\frac{2\pi}{3}} \overline{a}_{1,1} \Big(\widehat{L}_x+i\widehat{L}_y\Big)\nonumber\\
    &+i\sqrt{\frac{4\pi}{3}}\overline{a}_{1,0}\widehat{L}_z\nonumber\\
    &=i\sqrt{\frac{2\pi}{3}} \overline{a}_{1,-1} \widehat{L}_{-} -i\sqrt{\frac{2\pi}{3}} \overline{a}_{1,1} \widehat{L}_{+}+i\sqrt{\frac{4\pi}{3}}\overline{a}_{1,0}\widehat{L}_z,
    \label{eq:PetaL-app}
\end{align}
where $\widehat{L}_{\pm} = \widehat{L}_{x}\pm i \widehat{L}_{y}$ are the standard ladder operators. From the well-known angular momentum algebra, we also have the properties
\begin{align}
    &\widehat{L}_{z}Y_l^m=mY_l^m,\\
    &\widehat{L}_{\pm}Y_l^{m}=c_{lm}^{\pm}Y_l^{m\pm1}; c_{lm}^{\pm}\equiv \sqrt{l(l+1)-m(m\pm1)}. \label{eq: Lpm on Y}
\end{align}
Clearly, we have $c^+_{ll} = c^{-}_{l,-l}=0$. Furthermore, Eq.~\eqref{eq: Lpm on Y} implies that acting on $\eta$ in Eq.~\eqref{eq:eta_SH} by $\Big[\vec{\mathbb{P}}_\eta \times\vec{\mathbf{S}}\Big]\cdot \vec{\nabla}_\mathcal{S}$ couples only coefficients $a_{lm}$ with different $m$ but fixed $l$.

With the above background, we now focus on evaluating the third term; we have
\begin{align}
    &\mu\Big[~\vec{\mathbb{P}}_\eta \times \vec{\mathbf{S}}\Big]\cdot \vec{\nabla}_\mathcal{S}F_0\nonumber \\
    &=\mu\Big[i\vec{\mathbb{P}}_\eta \cdot  \widehat{\mathbf{L}}\Big]F_0 \nonumber\\
    &= \frac{1}{2\pi} \sum_{l=0}^\infty \sqrt{\pi(2l+1)}\Big[\beta~g_1(\omega) + (1-\beta)g_2(\omega)(-1)^l\Big] \nonumber\\
    &\times\frac{\mathcal{I}_l(\kappa)}{\mathcal{I}_0(\kappa)}\Big[i\vec{\mathbb{P}}_\eta \cdot  \widehat{\mathbf{L}}\Big]Y_l^0,
\end{align}
where we have used Eq.~\eqref{eq:F0-expansion-app}. Using Eq.~\eqref{eq:PetaL-app}, we next have
\begin{align}
   &\Big[i\vec{\mathbb{P}}_\eta \cdot  \widehat{\mathbf{L}}\Big]Y_l^0 \nonumber\\
   &= \bigg[i\sqrt{\frac{2\pi}{3}} \overline{a}_{1,-1} \widehat{L}_{-} -i\sqrt{\frac{2\pi}{3}} \overline{a}_{1,1} \widehat{L}_{+}+i\sqrt{\frac{4\pi}{3}}\overline{a}_{1,0}\widehat{L}_z\bigg]Y_l^0 \nonumber\\
   &= \bigg[i\sqrt{\frac{2\pi}{3}} \overline{a}_{1,-1} \widehat{L}_{-} -i\sqrt{\frac{2\pi}{3}} \overline{a}_{1,1} \widehat{L}_{+}\bigg]Y_l^0 \nonumber\\
   &= i\sqrt{\frac{2\pi}{3}} \overline{a}_{1,-1} c_{l0}^-Y_l^{-1} -i\sqrt{\frac{2\pi}{3}} \overline{a}_{1,1} c_{l0}^+Y_l^1 \nonumber\\
   &= i \sqrt{\frac{2\pi l(l+1)}{3}}\bigg[\overline{a}_{1,-1} Y_l^{-1}-\overline{a}_{1,1} Y_l^{1}\bigg].
\end{align}
This leads to the result that
\begin{align}
    &\mu\Big[~\vec{\mathbb{P}}_\eta \times \vec{\mathbf{S}}\Big]\cdot \vec{\nabla}_\mathcal{S}F_0 \nonumber\\
    &= \frac{\mu}{2\pi} \sum_{l=0}^\infty \sqrt{\pi(2l+1)}\Big[\beta~g_1(\omega) + (1-\beta)(-1)^lg_2(\omega)\Big] \nonumber\\
    &\times \frac{\mathcal{I}_l(\kappa)}{\mathcal{I}_0(\kappa)} i \sqrt{\frac{2\pi l(l+1)}{3}}\bigg[\overline{a}_{1,-1} Y_l^{-1}-\overline{a}_{1,1} Y_l^{1}\bigg] \nonumber\\
    &= i\mu \sum_{l=0}^\infty \sqrt{\frac{ l(l+1)(2l+1)}{6}}\Big[\beta~g_1(\omega) + (1-\beta)(-1)^lg_2(\omega)\Big] \nonumber\\
    &\times \frac{\mathcal{I}_l(\kappa)}{\mathcal{I}_0(\kappa)}  \bigg[\overline{a}_{1,-1} Y_l^{-1}-\overline{a}_{1,1} Y_l^{1}\bigg].
\end{align}
We can write the above result in a compact way by defining the function
\begin{align}
    &\Gamma_l(\omega,\beta,\kappa) \nonumber\\
    &\equiv \sqrt{\frac{ l(l+1)(2l+1)}{6}}\Big[\beta~g_1(\omega) + (1-\beta)(-1)^lg_2(\omega)\Big]\frac{\mathcal{I}_l(\kappa)}{\mathcal{I}_0(\kappa)},
\end{align}
yielding
\begin{align}
    &\mu\Big[~\vec{\mathbb{P}}_\eta \times \vec{\mathbf{S}}\Big]\cdot \vec{\nabla}_\mathcal{S}F_0 \nonumber\\
    &= -i\mu\sum_{l=0}^\infty \sum_{m=-l}^l \delta_{|m|,1}m \Gamma_l(\omega,\beta,\kappa) \overline{a}_{1,m}Y_l^m.
\end{align}

We now focus on the fourth term on the rhs of Eq.~\eqref{eq:eta_evolution}; proceeding as for the third term, we get
\begin{widetext}
\begin{align}
    &\mu\Big[~\vec{\mathbb{P}}_\eta \times \vec{\mathbf{S}}\Big]\cdot \vec{\nabla}_\mathcal{S}\eta \nonumber \\
    &=\mu\Big[i\vec{\mathbb{P}}_\eta \cdot  \widehat{\mathbf{L}}\Big]\eta\nonumber\\
    &=\mu \sum_{l=1}^\infty\sum_{m=-l}^la_{lm} \Big[i\vec{\mathbb{P}}_\eta \cdot  \widehat{\mathbf{L}}\Big]Y_l^m \nonumber\\
    &= i \mu\sqrt{\frac{2\pi}{3}}\sum_{l=1}^\infty\sum_{m=-l}^la_{lm}\bigg[\overline{a}_{1,-1} \widehat{L}_{-} - \overline{a}_{1,1} \widehat{L}_{+}+\sqrt{2}\overline{a}_{1,0}\widehat{L}_z\bigg]Y_l^m \nonumber\\
    &= i {\mu} \sqrt{\frac{2\pi}{3}}\sum_{l=1}^\infty\sum_{m=-l}^la_{lm}\bigg[\overline{a}_{1,-1} c_{lm}^{-}Y_l^{m-1} - \overline{a}_{1,1} c_{lm}^{+}Y_l^{m+1} +\sqrt{2}m\overline{a}_{1,0}Y_l^m\bigg].
\end{align}

Combining everything, we obtain from the continuity equation~\eqref{eq:eta_evolution} that
    \begin{align}
    \sum_{l=1}^\infty\sum_{m=-l}^l \frac{\partial a_{lm}}{\partial t}Y_l^m &= -\sum_{l=1}^\infty\sum_{m=-l}^l  \bigg[im\bigg\{\omega+\mu\big(2\beta-1\big) \frac{\mathcal{I}_1(\kappa)}{\mathcal{I}_0(\kappa)}\bigg\}a_{lm} -i \mu \delta_{|m|,1}m \Gamma_l(\omega,\beta,\kappa) \overline{a}_{1,m}\bigg] ~Y_l^m \nonumber\\
    &-i\mu \sqrt{\frac{2\pi}{3}}\sum_{l=1}^\infty\sum_{m=-l}^la_{lm}\bigg[\overline{a}_{1,-1} c_{lm}^{-}Y_l^{m-1} - \overline{a}_{1,1} c_{lm}^{+}Y_l^{m+1}+\sqrt{2}m\overline{a}_{1,0}Y_l^m\bigg]. \label{eq: now1}
\end{align}
Comparing the coefficient of $Y_l^m$s from both sides of Eq.~\eqref{eq: now1}, we obtain
    \begin{align}
    \frac{\partial a_{lm}}{\partial t} &= -  \bigg[im\bigg\{\omega+\mu\big(2\beta-1\big) \frac{\mathcal{I}_1(\kappa)}{\mathcal{I}_0(\kappa)}\bigg\}a_{lm} -i \mu \delta_{|m|,1}m \Gamma_l(\omega,\beta,\kappa) \overline{a}_{1,m}\bigg]  \nonumber\\
    &-i\mu \sqrt{\frac{2\pi}{3}}\bigg[\overline{a}_{1,-1} c_{l,m+1}^{-}a_{l,m+1} - \overline{a}_{1,1} c_{l,m-1}^{+}a_{l,m-1}+\sqrt{2}m\overline{a}_{1,0}a_{lm}\bigg]. \label{eq: now2}
\end{align}
Let us now focus on the evolution equation of $l=1$ modes. First, considering $l=1,~m=1$, we obtain, on comparing both sides of the above equation, that
\begin{equation}
        \frac{\partial a_{1,1}}{\partial t} = -i\bigg[\omega+\mu (2\beta-1)\frac{\mathcal{I}_1(\kappa)}{\mathcal{I}_0(\kappa)}\bigg]a_{1,1} +i\mu \bigg[\beta~g_1(\omega)-(1-\beta)g_2(\omega) \bigg]\frac{\mathcal{I}_1(\kappa)}{\mathcal{I}_0(\kappa)} \overline{a}_{1,1} - i\mu \sqrt{\frac{4\pi}{3}} \bigg[a_{1,1} \overline{a}_{1,0}-a_{1,0}\overline{a}_{1,1}\bigg].
\end{equation}
This is Eq.~\eqref{eq: a11 evolution} of the main text.

\section{Derivation of Eq.~\eqref{eq:eigenvalue}}
\label{app: Critical Point}
Using
\begin{align}
    a_{1,1} = \tilde{a}_{1,1}(\omega)e^{-i\Omega t},
\end{align}
and retaining only the linear terms in Eq.~\eqref{eq: a11 evolution}, we get
\begin{equation}
        -i\Omega \tilde{a}_{1,1}(\omega)e^{-i\Omega t} = -i\bigg[\omega+\mu(2\beta-1)\frac{\mathcal{I}_1(\kappa)}{\mathcal{I}_0(\kappa)}\bigg]\tilde{a}_{1,1}(\omega)e^{-i\Omega t} +i\mu \bigg[\beta~g_1(\omega)-(1-\beta)g_2(\omega) \bigg]\frac{\mathcal{I}_1(\kappa)}{\mathcal{I}_0(\kappa)} e^{-i\Omega t} \int_{-\infty}^\infty d\omega \tilde{a}_{1,1}(\omega).
    \end{equation}
Rearranging, we obtain
\begin{equation}
       \bigg[ \omega+ \mu (2\beta-1)\frac{\mathcal{I}_1(\kappa)}{\mathcal{I}_0(\kappa)}-\Omega\bigg]\tilde{a}_{1,1}(\omega) = \mu \bigg[\beta~g_1(\omega)-(1-\beta)g_2(\omega) \bigg]\frac{\mathcal{I}_1(\kappa)}{\mathcal{I}_0(\kappa)}  \int_{-\infty}^\infty d\omega \tilde{a}_{1,1}(\omega).
\end{equation}
Integrating both sides with respect to $\omega$, we get
\begin{align}
    \int_{-\infty}^\infty d\omega \tilde{a}_{1,1}(\omega) = \left[\int_{-\infty}^\infty d\omega \frac{\mu \bigg[\beta~g_1(\omega)-(1-\beta)g_2(\omega) \bigg]\frac{\mathcal{I}_1(\kappa)}{\mathcal{I}_0(\kappa)}}{\omega+ \mu (2\beta-1)\frac{\mathcal{I}_1(\kappa)}{\mathcal{I}_0(\kappa)}-\Omega} \right]\int_{-\infty}^\infty d\omega \tilde{a}_{1,1}(\omega),
\end{align}
which gives the dispersion relation determining the quantity $\Omega$ as 
\begin{equation}
    \bigg[\mu\frac{\mathcal{I}_1(\kappa)}{\mathcal{I}_0(\kappa)} \bigg]\int_{-\infty}^\infty d\omega \frac{ \bigg[\beta~g_1(\omega)-(1-\beta)g_2(\omega) \bigg]}{\omega+ \mu(2\beta-1) \frac{\mathcal{I}_1(\kappa)}{\mathcal{I}_0(\kappa)}-\Omega}=1.
\end{equation}
This gives Eq.~\eqref{eq:eigenvalue} in the main text.
\end{widetext}

\section{Synchronization Threshold for the Two-Lorentzian Spectrum}
\label{app:RT_two_lorentzian}
We derive the transverse synchronization threshold using the late-time transverse polarization order parameter $R_T$ for the
two-Lorentzian spectrum discussed in Sec.~\ref{sec:transverse_dephasing}. We define the normalized spectrum as
\begin{equation}
    \tilde g(\omega)\equiv
    \frac{g_1(\omega)-\alpha g_2(\omega)}
    {1-\alpha},
    \label{eq:geff_RT}
\end{equation}
where $g_1(\omega)$ and $g_2(\omega)$ are the two-Lorentzian distributions. This is related to the effective spectrum $h(\omega)$ defined in Eq.~\eqref{eq:h(omega)} through $h(\omega)=
    \bar\mu Q(\kappa)(1-\alpha)\,\tilde g(\omega)$

We use the self-consistency conditions in Eq.~(19) of
Ref.~\cite{Raffelt:2010za} to derive the threshold. In our normalization, the corresponding effective interaction strength is $Q(\kappa)(1-\alpha)\bar\mu$, and the two conditions become
\begin{align}
    \frac{\sin\delta}{Q(\kappa)(1-\alpha)\bar\mu_{\rm crit}}
    &=\pi\, \tilde g(\omega_r),    \label{eq:RT_threshold_imag}
    \\
    \frac{\cos\delta}{Q(\kappa)(1-\alpha)\bar\mu_{\rm crit}}
    &={\rm P.V.}\int_{-\infty}^{\infty}\frac{\tilde g(\omega)}{\omega-\omega_r}\,d\omega,
    \label{eq:RT_threshold_real}
\end{align}
where $\delta=2\theta_{\rm mix}$ and $\omega_r$ is a resonance frequency appearing in their derivation. For compactness, we define
\begin{equation}
    D_+=(\omega_r-\omega_0)^2+\sigma^2, \quad
    D_-=(\omega_r+\omega_0)^2+\sigma^2.
    \label{eq:Dpm_RT}
\end{equation}
The two terms on the r.h.s. of Eqs.~\eqref{eq:RT_threshold_imag} and
\eqref{eq:RT_threshold_real} are
\begin{equation}
    \pi\, \tilde g(\omega_r)=\frac{\sigma}{1-\alpha}\left(\frac{1}{D_+}-\frac{\alpha}{D_-}\right),
    \label{eq:h_resonance_RT}
\end{equation}
and
\begin{equation}
    {\rm P.V.}\int_{-\infty}^{\infty}\frac{\tilde g(\omega)}{\omega-\omega_r}\,d\omega = \frac{1}{1-\alpha}
    \left[\frac{\omega_0-\omega_r}{D_+}+ \alpha\frac{\omega_0+\omega_r}{D_-}\right].
    \label{eq:PV_RT}
\end{equation}
Dividing Eq.~\eqref{eq:RT_threshold_real} by
Eq.~\eqref{eq:RT_threshold_imag}, we obtain
\begin{equation}
    \frac{\omega_0-\omega_r}{D_+} + \alpha\frac{\omega_0+\omega_r}{D_-}=\sigma\cot\delta
    \left(\frac{1}{D_+}-\frac{\alpha}{D_-}\right).
    \label{eq:omega_r_RT}
\end{equation}
This gives a cubic equation for $\omega_r$. We choose the root
continuously connected to the $\alpha=0$ solution. Substituting
Eq.~\eqref{eq:h_resonance_RT} into
Eq.~\eqref{eq:RT_threshold_imag} gives
\begin{equation}
    \bar\mu_{\rm crit}=\frac{\sin\delta}{Q(\kappa)\sigma \left(D_+^{-1}-\alpha D_-^{-1}\right)},
    \label{eq:mucrit_RT}
\end{equation}
where $D_\pm$ are evaluated at the solution of
Eq.~\eqref{eq:omega_r_RT}. For $\alpha=0$, Eq.~\eqref{eq:omega_r_RT} gives
\begin{equation}
    \omega_r^{(0)}=\omega_0-\sigma\cot\delta,
    \label{eq:omega_r_alpha_zero_RT}
\end{equation}
and therefore
\begin{equation}
    D_+^{(0)}=\frac{\sigma^2}{\sin^2\delta}.
\end{equation}
The threshold at $\alpha=0$ reduces to
\begin{equation}
    \bar\mu_{\rm crit}=\frac{\sigma}{Q(\kappa)\sin\delta}.
    \label{eq:mucrit_alpha_zero_RT}
\end{equation}
For small $\alpha$, we expand
\begin{equation}
    \omega_r=\omega_r^{(0)}+\alpha\omega_r^{(1)}+{\cal O}(\alpha^2).
\end{equation}
Expanding Eq.~\eqref{eq:omega_r_RT} gives
\begin{equation}
    \omega_r^{(1)}=\frac{2\sigma^2\omega_0}{\sigma^2-4\sigma\omega_0\sin\delta\cos\delta+4\omega_0^2\sin^2\delta}.
\end{equation}
Using this result in Eq.~\eqref{eq:mucrit_RT}, we obtain
\begin{equation}
    \bar\mu_{\rm crit}=\frac{\sigma}{Q(\kappa)\sin\delta}\left[ 1+K_1(\delta)\alpha+{\cal O}(\alpha^2)\right],
    \label{eq:mucrit_small_alpha_RT}
\end{equation}
where
\begin{equation}
    K_1(\delta)=\frac{\sigma^2-4\sigma\omega_0\sin\delta\cos\delta}{\sigma^2-4\sigma\omega_0\sin\delta\cos\delta +4\omega_0^2\sin^2\delta}.
    \label{eq:K1_RT}
\end{equation}
In the small-tilt limit, $K_1(\delta)\to1$, giving
\begin{equation}
    \bar\mu_{\rm crit}\simeq\frac{\sigma}{Q(\kappa)\sin\delta}(1+\alpha), \quad (\alpha,\delta\ll1).
    \label{eq:mucrit_small_alpha_delta_RT}
\end{equation}
Finally, as $\delta\to0$, Eq.~\eqref{eq:omega_r_alpha_zero_RT} gives $\omega_r\simeq-\sigma\cot\delta$, so that $\bar\mu_{\rm crit}\propto1/\sin\delta$.


\bibliography{References}

@article{PhysRevLett.106.160601,
  title = {Diffusion with Stochastic Resetting},
  author = {Evans, Martin R. and Majumdar, Satya N.},
  journal = {Phys. Rev. Lett.},
  volume = {106},
  issue = {16},
  pages = {160601},
  numpages = {4},
  year = {2011},
  month = {Apr},
  publisher = {American Physical Society},
  doi = {10.1103/PhysRevLett.106.160601},
  url = {https://link.aps.org/doi/10.1103/PhysRevLett.106.160601}
}

@article{Evans2020,
  author  = {Evans, Martin R. and Majumdar, Satya N. and Schehr, Gr{\'e}gory},
  title   = {Stochastic Resetting and Applications},
  journal = {Journal of Physics A: Mathematical and Theoretical},
  volume  = {53},
  number  = {19},
  pages   = {193001},
  year    = {2020},
  doi     = {10.1088/1751-8121/ab7cfe}
}

@article{Bressloff2024,
  author  = {Bressloff, Paul C.},
  title   = {Global density equations for interacting particle systems with stochastic resetting: From overdamped Brownian motion to phase synchronization},
  journal = {Chaos},
  volume  = {34},
  number  = {4},
  pages   = {043101},
  year    = {2024},
  doi     = {10.1063/5.0196626}
}

@article{PhysRevE.109.064137,
  title = {Kuramoto model subject to subsystem resetting: How resetting a part of the system may synchronize the whole of it},
  author = {Majumder, Rupak and Chattopadhyay, Rohitashwa and Gupta, Shamik},
  journal = {Phys. Rev. E},
  volume = {109},
  issue = {6},
  pages = {064137},
  numpages = {22},
  year = {2024},
  month = {Jun},
  publisher = {American Physical Society},
  doi = {10.1103/PhysRevE.109.064137},
  url = {https://link.aps.org/doi/10.1103/PhysRevE.109.064137}
}

@article{np7q-hxld,
  title = {Manipulating Phases in Many-Body Interacting Systems with Subsystem Resetting},
  author = {Acharya, Anish and Majumder, Rupak and Gupta, Shamik},
  journal = {Phys. Rev. Lett.},
  volume = {135},
  issue = {12},
  pages = {127103},
  numpages = {10},
  year = {2025},
  month = {Sep},
  publisher = {American Physical Society},
  doi = {10.1103/np7q-hxld},
  url = {https://link.aps.org/doi/10.1103/np7q-hxld}
}

@article{d82g-qyvf,
  title = {Analytical approach to subsystem resetting in generalized Kuramoto models},
  author = {Majumder, Rupak and Acharya, Anish and Gupta, Shamik},
  journal = {Phys. Rev. E},
  volume = {114},
  issue = {2},
  pages = {024108},
  numpages = {24},
  year = {2026},
  month = {Aug},
  publisher = {American Physical Society},
  doi = {10.1103/d82g-qyvf},
  url = {https://link.aps.org/doi/10.1103/d82g-qyvf}
}

@article{PhysRevLett.77.3700,
  title = {Absence of Self-Averaging and Universal Fluctuations in Random Systems near Critical Points},
  author = {Aharony, Amnon and Harris, A. Brooks},
  journal = {Phys. Rev. Lett.},
  volume = {77},
  issue = {18},
  pages = {3700--3703},
  numpages = {0},
  year = {1996},
  month = {Oct},
  publisher = {American Physical Society},
  doi = {10.1103/PhysRevLett.77.3700},
  url = {https://link.aps.org/doi/10.1103/PhysRevLett.77.3700}
}

@article{PhysRevE.52.3469,
  title = {Lack of self-averaging in critical disordered systems},
  author = {Wiseman, Shai and Domany, Eytan},
  journal = {Phys. Rev. E},
  volume = {52},
  issue = {4},
  pages = {3469--3484},
  numpages = {0},
  year = {1995},
  month = {Oct},
  publisher = {American Physical Society},
  doi = {10.1103/PhysRevE.52.3469},
  url = {https://link.aps.org/doi/10.1103/PhysRevE.52.3469}
}

@misc{DLMF,
  author       = {Olver, Frank W. J. and Olde Daalhuis, Adri B. and
                  Lozier, Daniel W. and Schneider, Barry I. and
                  Boisvert, Ronald F. and Clark, Charles W. and
                  Miller, Bruce R. and Saunders, Bonita V. and
                  Cohl, Howard S. and McClain, Mark A.},
  title        = {NIST Digital Library of Mathematical Functions},
  howpublished = {\url{https://dlmf.nist.gov/}},
  year         = {2026},
  note         = {Release 1.2.7 of 2026-06-15},
}

@book{kuramoto1984chemical,
  author    = {Yoshiki Kuramoto},
  title     = {Chemical Oscillations, Waves, and Turbulence},
  series    = {Springer Series in Synergetics},
  publisher = {Springer},
  address   = {Berlin, Heidelberg},
  year      = {1984},
  isbn      = {978-3-642-69689-3}
}

@article{strogatz2000kuramoto,
  author    = {Steven H. Strogatz},
  title     = {From {K}uramoto to {C}rawford: {E}xploring the onset of synchronization in populations of coupled oscillators},
  journal   = {Physica D},
  volume    = {143},
  number    = {1--4},
  pages     = {1--20},
  year      = {2000},
  publisher = {Elsevier},
  doi       = {10.1016/S0167-2789(00)00094-4}
}

@article{acebron2005kuramoto,
  author    = {Juan A. Acebr{\'o}n and Luis L. Bonilla and Conrad J. P{\'e}rez Vicente and F{\'e}lix Ritort and Renato Spigler},
  title     = {The Kuramoto model: A simple paradigm for synchronization phenomena},
  journal   = {Reviews of Modern Physics},
  volume    = {77},
  number    = {1},
  pages     = {137--185},
  year      = {2005},
  publisher = {American Physical Society},
  doi       = {10.1103/RevModPhys.77.137}
}

@article{gupta2014kuramoto,
  author    = {Shamik Gupta and Alessandro Campa and Stefano Ruffo},
  title     = {Kuramoto model of synchronization: Equilibrium and nonequilibrium aspects},
  journal   = {Journal of Statistical Mechanics: Theory and Experiment},
  volume    = {2014},
  number    = {8},
  pages     = {R08001},
  year      = {2014},
  publisher = {IOP Publishing},
  doi       = {10.1088/1742-5468/2014/08/R08001}
}

@book{gupta2018statistical,
  author    = {Shamik Gupta and Alessandro Campa and Stefano Ruffo},
  title     = {Statistical Physics of Synchronization},
  publisher = {Springer},
  address   = {Berlin},
  year      = {2018},
  isbn      = {978-3-319-70760-8},
  doi       = {10.1007/978-3-319-70761-5}
}

@misc{majumder2025finitesizefluctuationsstochasticcoupled,
      title={Finite-size fluctuations for stochastic coupled oscillators: A general theory}, 
      author={Rupak Majumder and Julien Barré and Shamik Gupta},
      year={2025},
      eprint={2510.02448},
      archivePrefix={arXiv},
      primaryClass={cond-mat.stat-mech},
      url={https://arxiv.org/abs/2510.02448}, 
}

@article{chandra2019continuous,
  author    = {Sarthak Chandra and Michelle Girvan and Edward Ott},
  title     = {Continuous versus discontinuous transitions in the D-dimensional generalized Kuramoto model: Odd D is different},
  journal   = {Physical Review X},
  volume    = {9},
  number    = {1},
  pages     = {011002},
  year      = {2019},
  month     = jan,
  publisher = {American Physical Society},
  doi       = {10.1103/PhysRevX.9.011002}
}

@article{majumder2026synchronizationannealeddisorderhigherharmonic,
  title = {Synchronization with annealed disorder and higher-harmonic interactions in arbitrary dimensions: When two dimensions are special},
  author = {Majumder, Rupak and Gupta, Shamik},
  journal = {APS Open Sci.},
  volume = {1},
  pages = {000069},
  numpages = {57},
  year = {2026},
  month = {Jul},
  publisher = {American Physical Society},
  doi = {10.1103/cqwb-j1xy},
  url = {https://link.aps.org/doi/10.1103/cqwb-j1xy}
}

@article{Johns:2021qby,
    author = "Johns, Lucas",
    title = "{Collisional Flavor Instabilities of Supernova Neutrinos}",
    eprint = "2104.11369",
    archivePrefix = "arXiv",
    primaryClass = "hep-ph",
    doi = "10.1103/PhysRevLett.130.191001",
    journal = "Phys. Rev. Lett.",
    volume = "130",
    number = "19",
    pages = "191001",
    year = "2023"
}

@article{Richers:2021xtf,
    author = "Richers, Sherwood and Willcox, Donald and Ford, Nicole",
    title = "{Neutrino fast flavor instability in three dimensions}",
    eprint = "2109.08631",
    archivePrefix = "arXiv",
    primaryClass = "astro-ph.HE",
    reportNumber = "N3AS-21-013",
    doi = "10.1103/PhysRevD.104.103023",
    journal = "Phys. Rev. D",
    volume = "104",
    number = "10",
    pages = "103023",
    year = "2021"
}

@article{Fiorillo:2024pns,
    author = "Fiorillo, Damiano F. G. and Raffelt, Georg G.",
    title = "{Theory of neutrino slow flavor evolution. Part I. Homogeneous medium}",
    eprint = "2412.02747",
    archivePrefix = "arXiv",
    primaryClass = "hep-ph",
    doi = "10.1007/JHEP04(2025)146",
    journal = "JHEP",
    volume = "04",
    number = "04",
    pages = "146",
    year = "2025"
}

@article{Wu:2021uvt,
    author = "Wu, Meng-Ru and George, Manu and Lin, Chun-Yu and Xiong, Zewei",
    title = "{Collective fast neutrino flavor conversions in a 1D box: Initial conditions and long-term evolution}",
    eprint = "2108.09886",
    archivePrefix = "arXiv",
    primaryClass = "hep-ph",
    doi = "10.1103/PhysRevD.104.103003",
    journal = "Phys. Rev. D",
    volume = "104",
    number = "10",
    pages = "103003",
    year = "2021"
}

@article{Capozzi:2019lso,
    author = "Capozzi, Francesco and Raffelt, Georg and Stirner, Tobias",
    title = "{Fast Neutrino Flavor Conversion: Collective Motion vs. Decoherence}",
    eprint = "1906.08794",
    archivePrefix = "arXiv",
    primaryClass = "hep-ph",
    reportNumber = "MPP-2019-120",
    doi = "10.1088/1475-7516/2019/09/002",
    journal = "JCAP",
    volume = "09",
    number = "09",
    pages = "002",
    year = "2019"
}

@article{Duan:2010bg,
    author = "Duan, Huaiyu and Fuller, George M. and Qian, Yong-Zhong",
    title = "{Collective Neutrino Oscillations}",
    eprint = "1001.2799",
    archivePrefix = "arXiv",
    primaryClass = "hep-ph",
    reportNumber = "LA-UR-09-08309, INT-PUB-10-001",
    doi = "10.1146/annurev.nucl.012809.104524",
    journal = "Ann. Rev. Nucl. Part. Sci.",
    volume = "60",
    pages = "569--594",
    year = "2010"
}

@article{Mirizzi:2015eza,
      author         = "Mirizzi, Alessandro and Tamborra, Irene and Janka,
                        Hans-Thomas and Saviano, Ninetta and Scholberg, Kate and
                        Bollig, Robert and Hudepohl, Lorenz and Chakraborty,
                        Sovan",
      title          = "{Supernova Neutrinos: Production, Oscillations and
                        Detection}",
      journal        = "Riv. Nuovo Cim.",
      volume         = "39",
      year           = "2016",
      number         = "1-2",
      pages          = "1-112",
      doi            = "10.1393/ncr/i2016-10120-8",
      eprint         = "1508.00785",
      archivePrefix  = "arXiv",
      primaryClass   = "astro-ph.HE",
      SLACcitation   = "%%CITATION = ARXIV:1508.00785;%%"
}

@article{Martin:2019gxb,
    author = "Martin, Joshua D. and Yi, Changhao and Duan, Huaiyu",
    title = "{Dynamic fast flavor oscillation waves in dense neutrino gases}",
    eprint = "1909.05225",
    archivePrefix = "arXiv",
    primaryClass = "hep-ph",
    doi = "10.1016/j.physletb.2019.135088",
    journal = "Phys. Lett. B",
    volume = "800",
    pages = "135088",
    year = "2020"
}

@article{Sigl:1992fn,
    author = "Sigl, G. and Raffelt, G.",
    doi = "10.1016/0550-3213(93)90175-O",
    journal = "Nucl. Phys. B",
    pages = "423--451",
    reportNumber = "MPI-PH-92-112",
    title = "{General kinetic description of relativistic mixed neutrinos}",
    volume = "406",
    year = "1993"
}

@article{Pantaleone:1992eq,
    author = "Pantaleone, James T.",
    doi = "10.1016/0370-2693(92)91887-F",
    journal = "Phys. Lett. B",
    pages = "128--132",
    reportNumber = "DOE-ER-40561-056, INT-92-07-01",
    title = "{Neutrino oscillations at high densities}",
    volume = "287",
    year = "1992"
}

@article{Sawyer:2005jk,
    author = "Sawyer, R.F.",
    archivePrefix = "arXiv",
    doi = "10.1103/PhysRevD.72.045003",
    eprint = "hep-ph/0503013",
    journal = "Phys. Rev. D",
    pages = "045003",
    title = "{Speed-up of neutrino transformations in a supernova environment}",
    volume = "72",
    year = "2005"
}

@article{Duan:2005cp,
    author = "Duan, Huaiyu and Fuller, George M. and Qian, Yong-Zhong",
    archivePrefix = "arXiv",
    doi = "10.1103/PhysRevD.74.123004",
    eprint = "astro-ph/0511275",
    journal = "Phys. Rev. D",
    pages = "123004",
    title = "{Collective neutrino flavor transformation in supernovae}",
    volume = "74",
    year = "2006"
}

@article{Hannestad:2006nj,
    author = "Hannestad, Steen and Raffelt, Georg G. and Sigl, Guenter and Wong, Yvonne Y.Y.",
    archivePrefix = "arXiv",
    doi = "10.1103/PhysRevD.74.105010",
    eprint = "astro-ph/0608695",
    journal = "Phys. Rev. D",
    note = "[Erratum: Phys.Rev.D 76, 029901 (2007)]",
    pages = "105010",
    reportNumber = "MPP-2006-102",
    title = "{Self-induced conversion in dense neutrino gases: Pendulum in flavour space}",
    volume = "74",
    year = "2006"
}

@article{Raffelt:2007yz,
    author = "Raffelt, G.G. and Sigl, G.",
    archivePrefix = "arXiv",
    doi = "10.1103/PhysRevD.75.083002",
    eprint = "hep-ph/0701182",
    journal = "Phys. Rev. D",
    pages = "083002",
    reportNumber = "MPP-2007-6",
    title = "{Self-induced decoherence in dense neutrino gases}",
    volume = "75",
    year = "2007"
}

@article{Duan:2007mv,
    author = "Duan, Huaiyu and Fuller, George M. and Carlson, J. and Qian, Yong-Zhong",
    archivePrefix = "arXiv",
    doi = "10.1103/PhysRevD.75.125005",
    eprint = "astro-ph/0703776",
    journal = "Phys. Rev. D",
    pages = "125005",
    title = "{Analysis of Collective Neutrino Flavor Transformation in Supernovae}",
    volume = "75",
    year = "2007"
}

@article{EstebanPretel:2007ec,
    author = "Esteban-Pretel, Andreu and Pastor, Sergio and Tomas, Ricard and Raffelt, Georg G. and Sigl, Guenter",
    archivePrefix = "arXiv",
    doi = "10.1103/PhysRevD.76.125018",
    eprint = "0706.2498",
    journal = "Phys. Rev. D",
    pages = "125018",
    primaryClass = "astro-ph",
    reportNumber = "MPP-2007-60, IFIC-07-26",
    title = "{Decoherence in supernova neutrino transformations suppressed by deleptonization}",
    volume = "76",
    year = "2007"
}

@article{Bhattacharyya:2020dhu,
    author = "Bhattacharyya, Soumya and Dasgupta, Basudeb",
    title = "{Late-time behavior of fast neutrino oscillations}",
    eprint = "2005.00459",
    archivePrefix = "arXiv",
    primaryClass = "hep-ph",
    reportNumber = "TIFR/TH/20-15",
    doi = "10.1103/PhysRevD.102.063018",
    journal = "Phys. Rev. D",
    volume = "102",
    number = "6",
    pages = "063018",
    year = "2020"
}

@article{Bhattacharyya:2020jpj,
    author = "Bhattacharyya, Soumya and Dasgupta, Basudeb",
    title = "{Fast Flavor Depolarization of Supernova Neutrinos}",
    eprint = "2009.03337",
    archivePrefix = "arXiv",
    primaryClass = "hep-ph",
    reportNumber = "TIFR/TH/20-33",
    doi = "10.1103/PhysRevLett.126.061302",
    journal = "Phys. Rev. Lett.",
    volume = "126",
    number = "6",
    pages = "061302",
    year = "2021"
}

@article{Sawyer:2008zs,
    author = "Sawyer, R.F.",
    archivePrefix = "arXiv",
    doi = "10.1103/PhysRevD.79.105003",
    eprint = "0803.4319",
    journal = "Phys. Rev. D",
    pages = "105003",
    primaryClass = "astro-ph",
    title = "{The multi-angle instability in dense neutrino systems}",
    volume = "79",
    year = "2009"
}

@article{Dasgupta:2017oko,
    author = "Dasgupta, Basudeb and Sen, Manibrata",
    title = "{Fast Neutrino Flavor Conversion as Oscillations in a Quartic Potential}",
    eprint = "1709.08671",
    archivePrefix = "arXiv",
    primaryClass = "hep-ph",
    reportNumber = "TIFR-TH-17-34",
    doi = "10.1103/PhysRevD.97.023017",
    journal = "Phys. Rev. D",
    volume = "97",
    number = "2",
    pages = "023017",
    year = "2018"
}

@article{Dasgupta:2009mg,
    author = "Dasgupta, Basudeb and Dighe, Amol and Raffelt, Georg G. and Smirnov, Alexei Yu.",
    archivePrefix = "arXiv",
    doi = "10.1103/PhysRevLett.103.051105",
    eprint = "0904.3542",
    journal = "Phys. Rev. Lett.",
    pages = "051105",
    primaryClass = "hep-ph",
    reportNumber = "MPP-2009-33",
    title = "{Multiple Spectral Splits of Supernova Neutrinos}",
    volume = "103",
    year = "2009"
}

@article{Chakraborty:2016lct,
    author = "Chakraborty, Sovan and Hansen, Rasmus Sloth and Izaguirre, Ignacio and Raffelt, Georg",
    title = "{Self-induced neutrino flavor conversion without flavor mixing}",
    eprint = "1602.00698",
    archivePrefix = "arXiv",
    primaryClass = "hep-ph",
    doi = "10.1088/1475-7516/2016/03/042",
    journal = "JCAP",
    volume = "03",
    number = "03",
    pages = "042",
    year = "2016"
}

@article{Raffelt:2010za,
    author = "Raffelt, Georg G. and Tamborra, Irene",
    archivePrefix = "arXiv",
    doi = "10.1103/PhysRevD.82.125004",
    eprint = "1006.0002",
    journal = "Phys. Rev. D",
    pages = "125004",
    primaryClass = "hep-ph",
    reportNumber = "MPP-2010-55",
    title = "{Synchronization versus decoherence of neutrino oscillations at intermediate densities}",
    volume = "82",
    year = "2010"
}

@article{Banerjee:2011fj,
    author = "Banerjee, Arka and Dighe, Amol and Raffelt, Georg",
    archivePrefix = "arXiv",
    doi = "10.1103/PhysRevD.84.053013",
    eprint = "1107.2308",
    journal = "Phys. Rev. D",
    pages = "053013",
    primaryClass = "hep-ph",
    reportNumber = "MPP-2011-81, TIFR-TH-11-30",
    title = "{Linearized flavor-stability analysis of dense neutrino streams}",
    volume = "84",
    year = "2011"
}

@article{Chakraborty:2016yeg,
    author = "Chakraborty, Sovan and Hansen, Rasmus and Izaguirre, Ignacio and Raffelt, Georg",
    archivePrefix = "arXiv",
    doi = "10.1016/j.nuclphysb.2016.02.012",
    eprint = "1602.02766",
    journal = "Nucl. Phys. B",
    pages = "366--381",
    primaryClass = "hep-ph",
    title = "{Collective neutrino flavor conversion: Recent developments}",
    volume = "908",
    year = "2016"
}

@article{Izaguirre:2016gsx,
    author = "Izaguirre, Ignacio and Raffelt, Georg and Tamborra, Irene",
    archivePrefix = "arXiv",
    doi = "10.1103/PhysRevLett.118.021101",
    eprint = "1610.01612",
    journal = "Phys. Rev. Lett.",
    number = "2",
    pages = "021101",
    primaryClass = "hep-ph",
    reportNumber = "INT-PUB-16-023, MPP-2016-266",
    title = "{Fast Pairwise Conversion of Supernova Neutrinos: A Dispersion-Relation Approach}",
    volume = "118",
    year = "2017"
}

@article{Capozzi:2017gqd,
    author = "Capozzi, Francesco and Dasgupta, Basudeb and Lisi, Eligio and Marrone, Antonio and Mirizzi, Alessandro",
    archivePrefix = "arXiv",
    doi = "10.1103/PhysRevD.96.043016",
    eprint = "1706.03360",
    journal = "Phys. Rev. D",
    number = "4",
    pages = "043016",
    primaryClass = "hep-ph",
    reportNumber = "TIFR-TH-17-31",
    title = "{Fast flavor conversions of supernova neutrinos: Classifying instabilities via dispersion relations}",
    volume = "96",
    year = "2017"
}

@article{Johns:2020qsk,
    author = "Johns, Lucas and Nagakura, Hiroki and Fuller, George M. and Burrows, Adam",
    title = "{Fast oscillations, collisionless relaxation, and spurious evolution of supernova neutrino flavor}",
    eprint = "2009.09024",
    archivePrefix = "arXiv",
    primaryClass = "hep-ph",
    doi = "10.1103/PhysRevD.102.103017",
    journal = "Phys. Rev. D",
    volume = "102",
    number = "10",
    pages = "103017",
    year = "2020"
}

@article{Tamborra:2020cul,
    author = "Tamborra, Irene and Shalgar, Shashank",
    title = "{New Developments in Flavor Evolution of a Dense Neutrino Gas}",
    eprint = "2011.01948",
    archivePrefix = "arXiv",
    primaryClass = "astro-ph.HE",
    doi = "10.1146/annurev-nucl-102920-050505",
    journal = "Ann. Rev. Nucl. Part. Sci.",
    volume = "71",
    pages = "165--188",
    year = "2021"
}

@article{Morinaga:2021vmc,
    author = "Morinaga, Taiki",
    title = "{Fast neutrino flavor instability and neutrino flavor lepton number crossings}",
    eprint = "2103.15267",
    archivePrefix = "arXiv",
    primaryClass = "hep-ph",
    doi = "10.1103/PhysRevD.105.L101301",
    journal = "Phys. Rev. D",
    volume = "105",
    number = "10",
    pages = "L101301",
    year = "2022"
}

@article{Dasgupta:2021gfs,
    author = "Dasgupta, Basudeb",
    title = "{Collective Neutrino Flavor Instability Requires a Crossing}",
    eprint = "2110.00192",
    archivePrefix = "arXiv",
    primaryClass = "hep-ph",
    reportNumber = "TIFR/TH/21-15",
    doi = "10.1103/PhysRevLett.128.081102",
    journal = "Phys. Rev. Lett.",
    volume = "128",
    number = "8",
    pages = "081102",
    year = "2022"
}

@article{Airen:2018nvp,
    author = "Airen, Sagar and Capozzi, Francesco and Chakraborty, Sovan and Dasgupta, Basudeb and Raffelt, Georg and Stirner, Tobias",
    archivePrefix = "arXiv",
    doi = "10.1088/1475-7516/2018/12/019",
    eprint = "1809.09137",
    journal = "JCAP",
    pages = "019",
    primaryClass = "hep-ph",
    reportNumber = "MPP-2018-224, TIFR-TH-18-25",
    title = "{Normal-mode Analysis for Collective Neutrino Oscillations}",
    number = "12",
    year = "2018"
}

@article{Capozzi:2018clo,
    author = "Capozzi, Francesco and Dasgupta, Basudeb and Mirizzi, Alessandro and Sen, Manibrata and Sigl, Guenter",
    archivePrefix = "arXiv",
    doi = "10.1103/PhysRevLett.122.091101",
    eprint = "1808.06618",
    journal = "Phys. Rev. Lett.",
    number = "9",
    pages = "091101",
    primaryClass = "hep-ph",
    reportNumber = "MPP-2018-206, TIFR/TH/18-27",
    title = "{Collisional triggering of fast flavor conversions of supernova neutrinos}",
    volume = "122",
    year = "2019"
}

@article{Penrose1960,
    author = {Penrose, Oliver},
    title = {Electrostatic Instabilities of a Uniform Non-Maxwellian Plasma},
    journal = {The Physics of Fluids},
    volume = {3},
    number = {2},
    pages = {258-265},
    year = {1960},
    month = {03},
    issn = {0031-9171},
    doi = {10.1063/1.1706024},
    url = {https://doi.org/10.1063/1.1706024},
}

@article{Fiorillo:2024b,
    author = "Fiorillo, Damiano F. G. and Raffelt, Georg G.",
    title = "{Theory of neutrino fast flavor evolution. Part I. Linear response theory and stability conditions.}",
    eprint = "2406.06708",
    archivePrefix = "arXiv",
    primaryClass = "hep-ph",
    doi = "10.1007/JHEP08(2024)225",
    journal = "JHEP",
    volume = "08",
    number = "08",
    pages = "225",
    year = "2024"
}

@article{Xiong:2022zqz,
    author = "Xiong, Zewei and Johns, Lucas and Wu, Meng-Ru and Duan, Huaiyu",
    title = "{Collisional flavor instability in dense neutrino gases}",
    eprint = "2212.03750",
    archivePrefix = "arXiv",
    primaryClass = "hep-ph",
    doi = "10.1103/PhysRevD.108.083002",
    journal = "Phys. Rev. D",
    volume = "108",
    number = "8",
    pages = "083002",
    year = "2023"
}

@article{Padilla-Gay:2021haz,
    author = "Padilla-Gay, Ian and Tamborra, Irene and Raffelt, Georg G.",
    title = "{Neutrino Flavor Pendulum Reloaded: The Case of Fast Pairwise Conversion}",
    eprint = "2109.14627",
    archivePrefix = "arXiv",
    primaryClass = "astro-ph.HE",
    doi = "10.1103/PhysRevLett.128.121102",
    journal = "Phys. Rev. Lett.",
    volume = "128",
    number = "12",
    pages = "121102",
    year = "2022"
}

@article{Bhattacharyya:2022eed,
    author = "Bhattacharyya, Soumya and Dasgupta, Basudeb",
    title = "{Elaborating the ultimate fate of fast collective neutrino flavor oscillations}",
    eprint = "2205.05129",
    archivePrefix = "arXiv",
    primaryClass = "hep-ph",
    doi = "10.1103/PhysRevD.106.103039",
    journal = "Phys. Rev. D",
    volume = "106",
    number = "10",
    pages = "103039",
    year = "2022"
}

@article{Nagakura:2022kic,
    author = "Nagakura, Hiroki and Zaizen, Masamichi",
    title = "{Time-Dependent and Quasisteady Features of Fast Neutrino-Flavor Conversion}",
    eprint = "2206.04097",
    archivePrefix = "arXiv",
    primaryClass = "astro-ph.HE",
    doi = "10.1103/PhysRevLett.129.261101",
    journal = "Phys. Rev. Lett.",
    volume = "129",
    number = "26",
    pages = "261101",
    year = "2022"
}

@article{Volpe:2023met,
    author = "Volpe, M. Cristina",
    title = "{Neutrinos from dense environments: Flavor mechanisms, theoretical approaches, observations, and new directions}",
    eprint = "2301.11814",
    archivePrefix = "arXiv",
    primaryClass = "hep-ph",
    doi = "10.1103/RevModPhys.96.025004",
    journal = "Rev. Mod. Phys.",
    volume = "96",
    number = "2",
    pages = "025004",
    year = "2024"
}

@article{Johns:2025mlm,
    author = "Johns, Lucas and Richers, Sherwood and Wu, Meng-Ru",
    title = "{Neutrino Oscillations in Core-Collapse Supernovae and Neutron Star Mergers}",
    eprint = "2503.05959",
    archivePrefix = "arXiv",
    primaryClass = "astro-ph.HE",
    reportNumber = "LA-UR-25-21809",
    doi = "10.1146/annurev-nucl-121423-100853",
    journal = "Ann. Rev. Nucl. Part. Sci.",
    volume = "75",
    number = "1",
    pages = "399--423",
    year = "2025"
}

@article{Dasgupta:2025quc,
    author = "Dasgupta, Basudeb and Mukherjee, Dwaipayan",
    title = "{Sufficient and necessary conditions for collective neutrino instability: Fast, slow, and mixed}",
    eprint = "2505.03886",
    archivePrefix = "arXiv",
    primaryClass = "hep-ph",
    reportNumber = "TIFR/TH/25-12",
    doi = "10.1103/bqkr-rcwk",
    journal = "Phys. Rev. D",
    volume = "112",
    number = "12",
    pages = "123049",
    year = "2025"
}

@article{Fiorillo:2026lyz,
    author = "Fiorillo, Damiano F. G. and Raffelt, Georg G.",
    title = "{Ubiquitous flavor pendulum}",
    eprint = "2602.02655",
    archivePrefix = "arXiv",
    primaryClass = "hep-ph",
    doi = "10.1103/tkx8-jpj6",
    journal = "Phys. Rev. D",
    volume = "113",
    number = "12",
    pages = "123033",
    year = "2026"
}

@article{Liu:2025muc,
    author = "Liu, Jiabao and Johns, Lucas and Nagakura, Hiroki and Zaizen, Masamichi and Yamada, Shoichi",
    title = "{Dynamical equilibria of fast neutrino flavor conversion}",
    eprint = "2509.26418",
    archivePrefix = "arXiv",
    primaryClass = "astro-ph.HE",
    doi = "10.1103/ydrn-np36",
    journal = "Phys. Rev. D",
    volume = "114",
    number = "4",
    pages = "L041306",
    year = "2026"
}

@article{Goimil-Garcia:2025ozm,
    author = "Goimil-Garc{\'\i}a, Manuel and Tamborra, Irene",
    title = "{Steady state of fast-oscillating neutrinos in an inhomogeneous medium}",
    eprint = "2509.22805",
    archivePrefix = "arXiv",
    primaryClass = "astro-ph.HE",
    doi = "10.1103/gdg9-rzns",
    journal = "Phys. Rev. D",
    volume = "112",
    number = "10",
    pages = "103011",
    year = "2025"
}

\end{document}